\documentclass{JHEP3}
\usepackage{amsfonts,amsmath,epsf}
\usepackage{graphics,rotating} 

\newcommand\sect[1]{\section{#1}} 
\newcommand\void[1]       {}

\newcommand\be            {\begin{equation}}
\newcommand\bea           {\begin{equation}\begin{array}l\displaystyle}
\newcommand\bearll        {\begin{array}{ll}\displaystyle}
\newcommand\bearrl        {\begin{array}{rl}\displaystyle}
\newcommand\ee            {\end{equation}}
\newcommand\eear          {\end{array}}
\newcommand\enl           {\\[1em]\displaystyle}
\newcommand\etb           {& \displaystyle}
\newcommand\erf[1]        {\eqref{#1}}
\newcommand\labl[1]       {\label{#1}\ee}

\def\Bone{\hbox{1\!\!\!\!1}}
\newcommand\corr[1]       {\big\langle\,#1\,\big\rangle}
\newcommand\eps           {\varepsilon}
\newcommand\Hom           {\text{Hom}}
\newcommand\id            {{\rm id}}
\newcommand\inda          {\text{Ind}_A}

\newcommand\mc            {\mathcal}
\newcommand\one           {{\bf1}}
\newcommand\ol            {\overline}
\newcommand\oti           {\,{\otimes}\,}
\newcommand\ota           {\,{\otimes}\!{}_A\,}
\newcommand\rank          {\text{rank}}
\newcommand\rrangle       {\rangle\!\rangle}
\newcommand\X             {{\ensuremath{\,\omega}}}

\newcommand\Cb            {\mathbb{C}}
\newcommand\Rb            {\mathbb{R}}
\newcommand\Zb            {\mathbb{Z}}

\newcommand\Cc            {\mathcal{C}}

\newcommand\Nc            {\mathcal{N}}
\newcommand\Rc            {\mathcal{R}}
\newcommand\Tc            {\mathcal{T}}
\newcommand\Vc            {\mathcal{V}}

\newcommand\tV {{\text{Vir}}}

\renewcommand\vec[1]{{\vert{#1}\rangle}}
\newcommand\vecc[1]{{\vert\,#1\,\rrangle}}
\newcommand\cev[1]{{\langle{#1}\vert}}
\newcommand\vac{{\vec 0}}
\newcommand\cav{{\cev 0}}

\title{Reflection and Transmission for Conformal Defects}

\author{Thomas Quella\\ 
KdV Institute for Mathematics, University of Amsterdam,
Plantage Muidergracht 24,
1018 TV Amsterdam -- NL\\
E-mail: \email{tquella@science.uva.nl}}

\author{Ingo Runkel\\
Department of Mathematics, King's College London,
Strand, London WC2R\;2LS -- UK\\
E-mail: \email{Ingo.Runkel@kcl.ac.uk}}

\author{G\'erard M.T. Watts\\
Department of Mathematics, King's College London,
Strand, London WC2R\;2LS -- UK\\
E-mail: \email{Gerard.Watts@kcl.ac.uk}}

\abstract{We consider conformal defects joining 
two conformal field theories along a line.
We define two new quantities associated to such defects  
in terms of expectation values of the stress tensors 
and we propose them as measures of the reflectivity and
transmissivity of the defect.
Their properties are investigated and they are
computed in a number of examples.
We obtain a complete answer for all defects in the Ising model and
between certain pairs of minimal models.
In the case of two conformal field theories with an enhanced symmetry
we restrict ourselves to non-trivial defects that can be obtained
by a coset construction.}


\preprint{{\sf hep-th/0611296}\\
{\sf KCL-MTH-06-12}\\
{\sf NSF-KITP-06-110}}

\begin{document}

\sect{Introduction}

Just as a conformal boundary condition describes a universality class
of boundary critical behaviour in a two-dimensional quantum system, 
a conformal defect is a universality class of critical behaviour at
a one-dimensional junction of two such quantum systems. It is therefore
of some interest to understand the properties of conformal defects, and
consequently there exist numerous publications emphasising a variety of
different aspects. To name just a few, there has been considerable
effort to clarify the role of defects and impurities in concrete
applications in statistical models
\cite{Oshikawa:1996ww,Oshikawa:1996dj} (see also references therein),
quantum wires 
\cite{Wong:1994pa,Chamon:2003tz,Oshikawa:2005fh,Friedan:2005bz,Friedan:2005ca}
and even for domain walls in string theory \cite{Bachas:2001vj}.
Other works focused on general constructive methods
\cite{Petkova:2000ip,Petkova:2001ag,Chui:2001kw,Coquereaux:2001di,%
Quella:2002ct,Fuchs:2002cm}
or structural implications
\cite{Graham:2003nc,Frohlich:2004ef,Frohlich:2006ch}. 
Finally, there are also articles which have originally been
written in a different context but have implications for the study of
defect systems 
\cite{Fuchs:1999zi,Recknagel:2002qq,Quella:2002ns,Fredenhagen:2005an,%
Fredenhagen:2006qw}. 

In order to set the stage, 
let us consider a conformally invariant quantum system on the complex plane,
which is inhomogeneous in the sense that the theory on the upper half-plane
is described by a conformal field theory $\text{CFT}_1$, and 
on the lower half-plane by a conformal field theory $\text{CFT}_2$,
possibly with a different chiral symmetry or even a different value
of the central charge.
The two CFTs meet along the real line, which constitutes a defect where
the fields of the two theories can have discontinuities or divergences.
If the defect has the property that inside every correlator
\be
  \lim_{y \rightarrow 0} \Big( T^1(x{+}iy) - \ol T{}^1(x{+}iy)  \Big)
  = \lim_{y \rightarrow 0} \Big( T^2(x{-}iy) - \ol T{}^2(x{-}iy) \Big)
  \qquad \text{for all} ~~ x \in \Rb ~~,
\labl{eq:conf-defect-bc}
where $T^{1,2}$ and $\ol T{}^{1,2}$ are the holomorphic and
anti-holomorphic 
components of the stress tensor of $\text{CFT}_{1,2}$, then the defect
is called {\em conformal}.
There are two special solutions to the condition
\erf{eq:conf-defect-bc}. Firstly, the two sides of
\erf{eq:conf-defect-bc} can 
individually be zero. Then the real line is a conformal boundary to
$\text{CFT}_1$ and $\text{CFT}_2$ separately and the two theories are
decoupled; such conformal defects are called `totally reflective' or
`factorising'. Second, it can be that on the real line we have 
$T^1(x) = T^2(x)$ 
and 
$\ol T{}^1(x) = \ol T{}^2(x)$ so that the defect is invisible to
correlators of the stress tensor. Such defects are called `totally
transmissive' or `topological'. The latter type of conformal defect can
only exist if the central charges of the two theories coincide. 
The properties of topological defects in rational conformal field theories
have been studied in detail in 
\cite{Petkova:2000ip,Coquereaux:2001di,Petkova:2000dv,Graham:2003nc,
Frohlich:2004ef,Frohlich:2006ch}.

The classification of a complete set of conformal defects joining two
given CFTs is a very difficult problem, just as is that of finding all
conformal boundary conditions. However, even for Virasoro minimal
models, where all conformal boundary conditions can be constructed
\cite{Cardy:1989ir}, apart from a few exceptions (see sections
\ref{sec:Ising} and \ref{sec:minmod}), it is not known how to obtain
all conformal defects.  The only systematic investigation of conformal
defects in minimal models which extend beyond pure transmission or
reflection has been performed in the Ising case
\cite{Oshikawa:1996ww,Oshikawa:1996dj}. For a single free boson, 
on the other hand, 
just one specific one-parameter family of conformal defects has
been discussed in \cite{Bachas:2001vj}. The situation is slightly
better in models with an enhanced chiral symmetry such as WZW models
or coset theories. Here one can apply a (nested) coset construction
\cite{Quella:2002ct} (see also \cite{Quella:2002ns,Quella:2002fk} for
a more explicit treatment of groups and cosets) to systematically
reduce the symmetry preserved by the defect. A different class of
non-factorising defects for WZW and coset models is conjectured to
arise from `permutation-like' boundary conditions in product CFTs. So
far however, the analysis of this type of defects has either been
restricted to the semi-classical regime \cite{Fredenhagen:2005an} or
to models which allow one to use insights from topological conformal
field  theories
\cite{Fredenhagen:2006qw,Brunner:2005fv,Caviezel:2005th}. 

Given this variety of constructions it is helpful to have a simple
quantity at one's disposal which is relatively easy to compute and
contains some basic information about the conformal defect.  One such
quantity is the $g$-function \cite{Affleck:1991tk}, which can be
defined by relating the conformal defect to a conformal boundary
condition in the folded model (see section~\ref{sec:ref-trans}).  We
propose, also in section~\ref{sec:ref-trans}, two additional such
quantities $\Rc$ and $\Tc$ (related via $\Rc+\Tc=1$), defined in terms
of expectation values of the stress tensors of $\text{CFT}_1$ and
$\text{CFT}_2$. Their properties suggest they might be useful quantitative  
indicators of the reflectivity and transmissivity of the conformal  
defect. This is motivated by the fact that $\Rc = 1$
for totally reflective defects and $\Rc=0$ for totally transmissive
defects, as well as $\Rc+\Tc=1$.  In
sections~\ref{sec:freebos}--\ref{sec:minmod} we proceed to compute $\Rc$ and
$\Tc$ for a selection of conformal defects in certain CFTs.  The
explicit models we consider are the free boson, the Ising model,
defects between WZW theories arising from the nested coset
construction, and those pairs of Virasoro minimal models which have a
product that is again a minimal model.  For two general rational CFTs
we describe, in section~\ref{sec:SubSym}, defects that are
transmissive only with respect to a common rational sub-symmetry
(which does not necessarily contain the Virasoro algebras of the two CFTs).
Altogether, we find that in the unitary examples treated, $\Rc$ and
$\Tc$ take values in the interval $[0,1]$, while in non-unitary
theories they can violate these bounds.  The bulk of the technical
computations has been gathered in several appendices.

\sect{Reflection and transmission coefficients}\label{sec:ref-trans}

\subsection{Definition of $\Rc$ and $\Tc$}

While it is possible to describe conformal defects as operators
between the Hilbert space of one CFT and another, it is more usual to
treat them as boundary conditions in an enlarged theory obtained by
`folding' the lower half-plane to lie on top of the upper half-plane
to give the product theory $\text{CFT}_1 \times \ol{\text{CFT}}_2$ on
the upper half-plane alone \cite{Wong:1994pa,Oshikawa:1996ww}.  Here
$\ol{\text{CFT}}_2$ stands for the theory obtained by exchanging
holomorphic and anti-holomorphic degrees of freedom in
$\text{CFT}_2$.\footnote{ Passing from $\text{CFT}_2$ to
  $\ol{\text{CFT}}_2$ may involve a choice of convention; the
  identification of degrees of freedom is only unique up to
  automorphisms of $\text{CFT}_2$. Once such an identification is
  fixed, the relation between conformal defects linking $\text{CFT}_1$
  to $\text{CFT}_2$ and conformal boundary conditions of $\text{CFT}_1
  \times \ol{\text{CFT}}_2$ is uniquely fixed as well.  }  The real
line then is a boundary for $\text{CFT}_1 \times \ol{\text{CFT}}_2$
and condition \erf{eq:conf-defect-bc} amounts to demanding the
boundary condition to be conformal in the sense of
\cite{Cardy:1984bb}.  In the folded picture, the topological defects
are a special case of so-called `permutation branes' studied in
\cite{Recknagel:2002qq}, and factorising defects correspond to
boundary conditions in the product theory, for which the boundary
state can be written as a product of boundary states for the
individual CFTs.

If we map the upper half-plane in the folded model to the exterior of
the unit circle then the boundary is represented by a `boundary state'
$\vec{b}$ in the bulk Hilbert space. This space is the tensor product
of the Hilbert spaces of $\text{CFT}_1$ and $\ol{\text{CFT}}_2$ and
the condition \erf{eq:conf-defect-bc} is
\be
\left(  (L^1_m + L^2_m) - (\ol L{}^1_{-m} + \ol L{}^2_{-m})
\right)  \vec b = 0\;.
\labl{eq:conf-def-bc}
It guarantees the existence of an infinite-dimensional conformal
symmetry in the defect system. 

Conformal boundary conditions (or defects) can be thought of as
special cases of integrable boundary conditions (or
defects)\
\cite{Fring:1993mp,Ghoshal:1993tm,Delfino:1994nx,Delfino:1994nr,%
Konik:1997gx,LeClair:1997gz,Bowcock:2004my}. 
Integrable boundary conditions have been mainly considered
for massive integrable field theories with a multiparticle
spectrum. For such theories an integrable boundary condition (or
defect) is almost entirely characterised by a reflection (or
reflection/transmission) matrix giving the amplitudes for a single
particle hitting the boundary (or defect) to emerge as a particle of a
different species. We would like to find a quantity which captures at
least some of the information contained in such a matrix in the case
of conformal field theories where the particle interpretation is
rather involved or missing.  

Let us first consider the case of free massless fields. To quote one
formula for the free boson (this is reviewed in section 3), the
boundary state representing the conformal defect satisfies 
\begin{equation}
 \left( a^i_m - S_{ij}\bar a^j_{-m} \right) \vec b = 0
\;,
\label{eq:fb}
\end{equation}
so that the reflection and transmission amplitudes for the bosonic
modes $a^i_m$ and $\bar a^i_{-m}$ are constant, given by the
matrix $S_{ij}$. 

We would like to extend this picture to more general conformal field
theories in which a particle interpretation is unknown or complicated.
In a general conformal field theory the only tool we have is the
Virasoro algebra and consequently it would be good if we could find an
analogue of \erf{eq:fb} involving the Virasoro algebra.  If we suppose
that
\begin{equation}
 \left( L^i_m - S_{ij}\ol L{}^j_{-m} \right) \vec b = 0
\;,
\label{eq:v}
\end{equation}
then we find that this can only be consistent 
with the Virasoro algebra 
for the choices
\be
S = \begin{pmatrix} 1 & 0 \\ 0 & 1 \end{pmatrix}
\;,\;\;
\hbox{ or }\;\;\;
S = \begin{pmatrix} 0 & 1  \\ 1 & 0 \end{pmatrix}
\;,
\ee
that is for purely reflecting or purely transmitting defects.

Since we cannot define a matrix $S_{ij}$ by \erf{eq:v} for a general
conformal defect, we consider instead the matrix
\be
  R_{ij}
= \frac{ \cev 0 L^i_2 \ol L{}^j_2 \vec b }
       { \cev 0 b \rangle }
\;.
\labl{eq:rijdef}
This would be simply related to the matrix $S_{ij}$ if \erf{eq:v}
held. Using general properties of boundary states, one can show that
the form of $R_{ij}$ is fixed up to a single parameter $\X_b$,
\be
  R = 
\frac{ c_1 c_2}{2(c_1 + c_2)}
\left[
\begin{pmatrix} \frac{c_1}{c_2} & 1 \\ 1 & \frac{c_2}{c_1}
\end{pmatrix}
+ \X_b
\begin{pmatrix} 1 & -1 \\ -1 & 1
\end{pmatrix}
\right]\;.
\labl{eq:R-via-omb}
The proof of this is given in appendix \ref{app:RT-omega}.
Note that considering $\cev 0 L^i_n \ol L{}^j_n \vec b$ in addition to $R_{ij}$
does not give more information. This follows since evaluating
$\langle 0 | L^i_n \ol L{}^j_{n+1}$ on \erf{eq:conf-def-bc} with $m=1$, one obtains a
simple recursion relation which leads to
\be
 \frac{ \cev 0 L^i_n \ol L{}^j_n \vec b }
       { \cev 0 b \rangle }
  = \frac{n(n^2-1)}{6} \, R_{ij}
  \qquad \text{for} ~~ n \ge 0 
  \;.
\ee

The matrix $R_{ij}$ is closely related to the `entropic admittance matrix' $Y_S(\omega)_{ij}$ introduced in \cite{Friedan:2005bz,Friedan:2005ca} 
to describe the entropy flow in junctions of quantum wires. The explicit relation
is obtained by Fourier transform of \cite[eqn.\,(87)]{Friedan:2005ca},\footnote{
  In \cite{Friedan:2005ca} the focus is on near-critical junctions in 
  bulk-critical quantum wires. We are concerned only with
  conformal defects, which correspond to critical junctions. 
  While \erf{eq:YS-def} obeys properties A,B,D,E,F listed in 
  \cite[sect.\,V]{Friedan:2005ca}, we are not certain how to 
  reconcile it with property C.}
\be
  Y_S(\omega)_{ij} =  
  f(\omega) \big( R_{ij} - \tfrac{c_i}{2} \delta_{ij} \big) = 
  f(\omega) \frac{c_1 \, c_2 \, (\omega_b{-}1) }{2\,(c_1{+}c_2)} 
  \begin{pmatrix} 1 & -1 \\ -1 & 1 \end{pmatrix} ~.
\labl{eq:YS-def}
with
$f(\omega) = \big( \beta / (2 \pi) \big)^3 \,
  \big(k \hbar v^2\big)^2 \,
  \big( 1 + ( \hbar \beta \omega / (2 \pi))^2 \big)/6$
and $\X_b$ the parameter appearing in \erf{eq:R-via-omb}.

Rather than characterising a defect by the value of $\X_b$, we
instead propose the following two
quantities which have very
appealing properties:
\bea
  \Rc = \frac{2}{c_1 + c_2}(R_{11} + R_{22})
  = \frac{c_1^{\,2} + 2 c_1 c_2 \X_b + c_2^{\,2}}{(c_1{+}c_2)^2} 
  \;,
  \enl
  \Tc = \frac{2}{c_1 + c_2}(R_{12} + R_{21})
  = \frac{2c_1 c_2 ( 1-\X_b)}{(c_1{+}c_2)^2} 
  \;.
\eear
\labl{eq:RT-via-omega}
They satisfy $\Rc=1$ for purely reflecting defects and $\Rc=0$ for
purely transmitting defects, and together with the obvious relation
$\Rc + \Tc = 1$ this prompts our calling them reflection and
transmission coefficients. They also have obvious physical
interpretations in the case of certain
defects in a single free boson or
free fermion theory, where $\Rc$ and $\Tc$ are the {\em probabilities}
of reflection or transmission of the free field modes.

The above definition of $\Rc$ and $\Tc$ is in terms of boundary states
in the `folded' theory.  We can also define them via the expectation
values of the components of the stress-energy tensor on the two sides
of the defect.  Consider the complex plane with a conformal defect on
the real line, and denote by $T^1$, $\ol T{}^1$ and $T^2$, $\ol T{}^2$
the components of the stress tensor of the CFTs on the two sides of
the defect.  Then instead of \erf{eq:RT-via-omega} we can write
\be
  \Rc = \frac{ \big\langle T^1 \ol T{}^1 + T^2 \ol T{}^2 
  \big\rangle_{1|2}}{\big\langle (T^1+ \ol T{}^2) (\ol T{}^1+T^2) 
  \big\rangle_{1|2}}
  \qquad \text{and} \qquad
  \Tc = \frac{ \big\langle T^1 T^2 + \ol T{}^1 \ol T{}^2 
  \big\rangle_{1|2}}{\big\langle (T^1+ \ol T{}^2) (\ol T{}^1+T^2) 
  \big\rangle_{1|2}} ~~.
\labl{eq:R-T-def}
Here, $T^1$ and $\ol T{}^1$ are inserted at the point $iy$ on the
upper half-plane, while $T^2$ and $\ol T{}^2$ are inserted at the
point ${-}iy$.  Since the numerator and the denominator in these
formulas are both proportional to $y^{-4}$, $\Rc$ and $\Tc$ do not
depend on the choice of $y$.
Note also that, as opposed to the $g$-function, 
because $\Rc$ and $\Tc$ are defined as quotients
they are not additive if one considers superpositions of defects.

\subsection{$\Rc$, $\Tc$ and topological defects}\label{sec:RT-top}

One very useful property of the quantities $\Rc$ and $\Tc$ is their
invariance under the action of topological defects, which we will now
explain.

First note that since the stress tensor is continuous across a
topological defect, the defect commutes with local conformal
transformations and can be deformed continuously without affecting the
value of a correlator.
This is the reason for the qualifier
`topological' (introduced in \cite{Bachas:2004sy}). Now consider two
topological defect lines $X$ and $Y$ which are running parallel to
each other. Moving them very close together, they look like a new
topological defect, which is called the {\em fused} defect $X \star
Y$. Altogether this defines the fusion ring of topological defects
\cite{Petkova:2000ip,Petkova:2001ag,Chui:2001kw,Coquereaux:2001di}. 
Let us call a topological defect {\em elementary} if it cannot be
written as the sum of two other topological defects. 
Then, even if one starts with two elementary topological defects $X$,
$Y$, the fused defect $X \star Y$ is typically no longer elementary.
One can also consider a topological defect $X$ close to a conformal
boundary $B$.  Since $X$ commutes with the stress tensor, moving the
defect against the boundary gives rise to a new conformal boundary
condition $X \star B$
\cite{Petkova:2001ag,Graham:2003nc,Frohlich:2004rj}.  This defines an
action of topological defects on boundary conditions.  Again, even if
$X$ and $B$ are elementary, $X \star B$ is typically not.\footnote{We
  will use the qualifier `elementary' for boundary conditions and
  conformal defects in the same sense as for topological defects.}

In the case where we have $\text{CFT}_1$ on the upper half-plane and
$\text{CFT}_2$ on the lower half-plane, separated by a conformal
defect $D$, we can place a topological defect $X$ of $\text{CFT}_1$ on
the line $\Rb + i L$, for some $L>0$, and a topological defect $Y$ of
$\text{CFT}_2$ on the line $\Rb - i L$.  In the limit $L\rightarrow 0$
we obtain a new conformal defect $X \star D \star Y$ (which is in
general not elementary even if $X,Y$ and $D$ were). As for conformal
boundary conditions, in this way we obtain an action of topological
defects on conformal defects.
 
One can also wonder if it is possible to fuse two parallel conformal defects
which are not topological. In this case the correlator does depend
on their distance and, much like in the OPE of two fields, one would
expect divergences as one takes the distance to zero (see
\cite{Bachas:2001vj} for an explicit calculation). We will not investigate
this situation in the present paper.
\medskip

Consider now 
the correlator \erf{eq:R-T-def} defining $\Rc$, but with topological
defects $X$ and $Y$ placed on the lines $\Rb \pm iL$.  As the defects
are topological, the resulting correlators will not depend on $L$.
Taking $L$ to infinity removes the defects from the complex plane,
multiplying the correlator by an overall constant which cancels
between the numerator and denominator, and we obtain the quantity $\Rc
\equiv \Rc(D)$ in the presence of the conformal defect $D$.  This is
nothing but the procedure of `inflating a defect in a world sheet'
used extensively in \cite{Frohlich:2004ef,Frohlich:2006ch}.  Taking
$L$ to zero, which we can do because $X$ and $Y$ commute with the
stress tensors, gives rise to the fused conformal defect $X \star D
\star Y$.  In this way we obtain the identity
\be
  \Rc(D) = \Rc(X \star D \star Y)
  \qquad \text{for~all~topological~defects~}X,Y ~.
\labl{eq:RD=RXDY}
Of course, the same holds also for $\Tc$. 
In other words, $\Rc$ and $\Tc$ are functions on the set of conformal
defects that are invariant under the action of topological defects.

Note that one can also define a `universal ground state degeneracy'
$g(D)$ for a conformal defect $D$ by considering the corresponding
quantity of the conformal boundary condition in the folded model as
defined in \cite{Affleck:1991tk}. However, unless acting with
so-called group-like defects (which are topological defects that have
an inverse w.r.t.\ defect fusion), $g(X \star D \star Y)$ will be
different from $g(D)$.

\medskip

Let $D$ be an elementary conformal defect. We say that another
elementary conformal defect $D'$ is {\em generated from} $D$ {\em by
  the action of topological defects} if there are topological defects
$X,Y$ such that $D'$ occurs in the decomposition of $X \star D \star
Y$ into elementary defects.  While \erf{eq:RD=RXDY} tells us that 
a weighted average of $\Rc(D')$ over all $D'$ that occur in the
decomposition of 
$X \star D \star Y$ is equal to $\Rc(D)$, under certain conditions we
even have $\Rc(D)= \Rc(D')$ for all such $D'$, as we will now see.

For an elementary conformal defect $D$, let $F = X \star D \star Y$
have the decomposition $F = \sum_\alpha D_\alpha$ into elementary
defects $D_\alpha$. We will work in the folded picture, and denote the
conformal boundary conditions corresponding to $D$, $F$, and
$D_\alpha$ by the same symbols. The topological defects $X$ and $Y$
can be expressed as a single topological defect $X\ol Y$ in the
product theory $\text{CFT}_1 \times \ol{\text{CFT}}_2$.

First note that by moving only a portion of the topological defect
$X\ol Y$ to a boundary with boundary condition $D$ we obtain the
identity 
\be
  \raisebox{-25pt}{
  \begin{picture}(300,50)
    \put(0,5){ \scalebox{.8}{\includegraphics{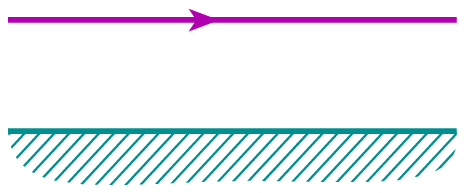}} }
    \put(0,5){
     \setlength{\unitlength}{.8pt}\put(-64,-504){
     \put(170,539)  {\scriptsize$ X \ol Y $}
     \put( 87,523)  {\scriptsize$ D $}
     }\setlength{\unitlength}{1pt}}
    \put(130,25){$=$}
    \put(150,5){ \scalebox{.8}{\includegraphics{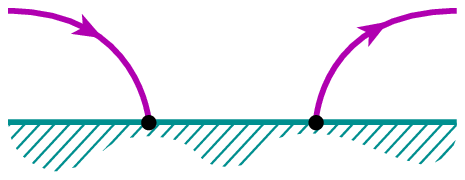}} }
    \put(150,5){
     \setlength{\unitlength}{.8pt}\put(-64,-504){
     \put( 80,554)  {\scriptsize$ X \ol Y $}
     \put(169,554)  {\scriptsize$ X \ol Y $}
     \put( 87,523)  {\scriptsize$ D $}
     \put(124,523)  {\scriptsize$ F $}
     \put(180,523)  {\scriptsize$ D $}
     \put(102,506)  {\scriptsize$ \ol\gamma(x) $}
     \put(149,506)  {\scriptsize$ \gamma(y) $}
     }\setlength{\unitlength}{1pt}}
  \end{picture}}
\labl{eq:fuse-part}
The portion of the defect $X\ol Y$ moved to the boundary fuses with
the boundary $D$ to become the boundary condition $F$ and the $X\ol
Y$ defect now ends and starts at the junctions of the boundary
conditions $D$ and $F$;
$\gamma(x)$ and $\ol\gamma(y)$ are the
(Virasoro-primary weight zero) boundary fields that mark
the end- and starting-points of the defect $X\ol Y$. 

Second, note that the fact that we can decompose $F = \sum_\alpha
D_\alpha$ means that we can find Virasoro-primary weight zero boundary
fields $P_\alpha$ on $F$ which form a complete set of orthogonal
idempotents w.r.t.\ to the OPE, i.e.\ $P_\alpha(x) P_\beta(y) =
\delta_{\alpha,\beta} P_\alpha(y)$ and $\sum_\alpha P_\alpha(x) =
\one_F$, the identity field on $F$.  In fact, the $P_\alpha(x)$ are
just the identity fields for the individual boundary conditions
$D_\alpha$.  For example, a correlator of some bulk fields on a disc
with boundary condition $D_\alpha$ is equal to a disc correlator with
the same bulk fields, but with boundary condition $F$ and an insertion
of $P_\alpha$ on the boundary.

Let now $\Phi$ be a bulk field of 
$\text{CFT}_1 \times \ol{\text{CFT}}_2$ that commutes with topological
defects of the form $X \ol Y$. We have the following equalities of
disc correlators, 
\be
  \corr{\Phi(0)}_{\text{disc}}^{D_\alpha}
  = 
  \raisebox{-30pt}{
  \begin{picture}(67,60)
    \put(-5,0){ \scalebox{.7}{\includegraphics{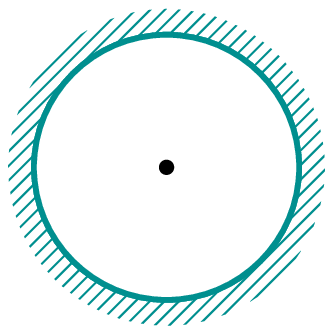}} }
    \put(-5,0){
     \setlength{\unitlength}{.7pt}\put(-8,-14){
     \put( 59, 49)  {\scriptsize$ \Phi $}
     \put( 44, 84)  {\scriptsize$ D_\alpha $}
     }\setlength{\unitlength}{1pt}}
  \end{picture}}
  \overset{(1)}{=} 
  \raisebox{-30pt}{
  \begin{picture}(70,60)
    \put(-5,0){ \scalebox{.7}{\includegraphics{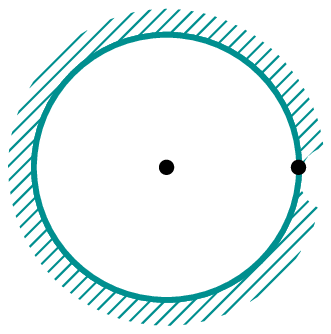}} }
    \put(-5,0){
     \setlength{\unitlength}{.7pt}\put(-8,-14){
     \put( 59, 49)  {\scriptsize$ \Phi $}
     \put( 44, 84)  {\scriptsize$ F $}
     \put(100, 54)  {\scriptsize$ P_\alpha $}
     }\setlength{\unitlength}{1pt}}
  \end{picture}}
  \overset{(2)}{=}
  \raisebox{-30pt}{
  \begin{picture}(70,60)
    \put(-5,0){ \scalebox{.7}{\includegraphics{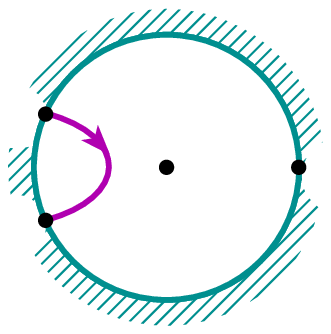}} }
    \put(-5,0){
     \setlength{\unitlength}{.7pt}\put(-8,-14){
     \put( 59, 49)  {\scriptsize$ \Phi $}
     \put( 42, 69)  {\scriptsize$ X\ol Y $}
     \put( 44, 84)  {\scriptsize$ F $}
     \put( 24, 57)  {\scriptsize$ D $}
     \put( 40, 30)  {\scriptsize$ F $}
     \put(100, 54)  {\scriptsize$ P_\alpha $}
     \put( 13, 73)  {\scriptsize$ \gamma $}
     \put( 14, 40)  {\scriptsize$ \ol\gamma $}
     }\setlength{\unitlength}{1pt}}
  \end{picture}}
  \overset{(3)}{=} 
  \raisebox{-30pt}{
  \begin{picture}(65,60)
    \put(-5,0){ \scalebox{.7}{\includegraphics{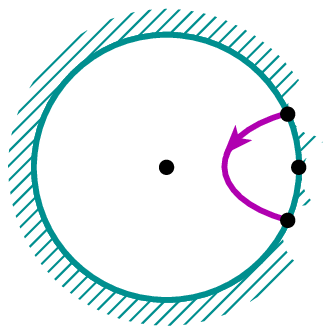}} }
    \put(-5,0){
     \setlength{\unitlength}{.7pt}\put(-8,-14){
     \put( 59, 49)  {\scriptsize$ \Phi $}
     \put( 65, 74)  {\scriptsize$ X\ol Y $}
     \put( 24, 57)  {\scriptsize$ D $}
     \put(100, 54)  {\scriptsize$ P_\alpha $}
     \put( 98, 75)  {\scriptsize$ \gamma $}
     \put( 99, 36)  {\scriptsize$ \ol\gamma $}
     }\setlength{\unitlength}{1pt}}
  \end{picture}}
\labl{eq:remove-defect}
Here in step (1) we replaced the boundary condition $D_\alpha$ by $F$
and an insertion of $P_\alpha$, in step (2) we used the inverse
transformation of \erf{eq:fuse-part}, and in (3) we moved $\gamma$ and
$\ol \gamma$ close to the $P_\alpha$ insertion and took the
topological defect $X\ol Y$ past the bulk field $\Phi$, as we can do
by assumption.  Suppose now that in addition the following two
conditions hold:
\\[.5em]
(i) Up to scalar multiples, there is a unique Virasoro-highest weight
boundary field of weight zero on the $D$ boundary (namely the identity
field $\one_D$).
\\[.5em]
(ii) The disc partition function with boundary condition $D_\alpha$,
$\langle \one \rangle_{\text{disc}}^{D_\alpha}$, is nonzero.
\\[.5em]
Then we can conclude further that \be
\corr{\Phi(0)}_{\text{disc}}^{D_\alpha} = C_{\alpha,\gamma,\ol\gamma}
\, \corr{\Phi(0)}_{\text{disc}}^{D} \labl{eq:RD-D'-aux1} for some
nonzero constant $C_{\alpha,\gamma,\ol\gamma}$ which depends on the
choice of $\gamma$, $\ol\gamma$ and $\alpha$, but not on the bulk
field $\Phi$. This has to be true, since the field one obtains by
collapsing the remaining defect bubble in \erf{eq:remove-defect} has
to be Virasoro-primary of weight zero, and by assumption (i) it thus
is proportional to the identity field $\one_D$; the constant
$C_{\alpha,\gamma,\ol\gamma}$ appearing in \erf{eq:RD-D'-aux1} is
nothing but this proportionality constant. It is nonzero because by
assumption (ii) the left hand side of \erf{eq:RD-D'-aux1} is nonzero
in the case $\Phi=\one$.

If we insert the identity \erf{eq:RD-D'-aux1} into the definition 
\erf{eq:RT-via-omega} of $\Rc$ (transformed to the unit disc), we obtain,
with the choices $\Phi=T^1\bar T^1 + T^2\bar T^2$
and $\Phi=\one$, respectively,
\be
  \Rc(D_\alpha) = 
  \frac{2}{c_1{+}c_2}\,
  \frac{\corr{ T^1 \ol T{}^1 + T^2 \ol T{}^2}_\text{disc}^{D_\alpha} }
       {\langle \one \rangle_\text{disc}^{D_\alpha} }
  =
  \frac{2}{c_1{+}c_2}\,\frac{
  C_{\alpha,\gamma,\ol\gamma}\,
  \corr{ T^1 \ol T{}^1 + T^2 \ol T{}^2}_\text{disc}^{D} }{
  C_{\alpha,\gamma,\ol\gamma}\,
  \langle \one \rangle_\text{disc}^{D} }
  = \Rc(D) ~.
\ee

Let us make two comments. First, our assumption that $D$ is an
elementary defect is necessary but not sufficient for condition (i) to
hold. Second, the requirement (ii) is not really a restriction,
because it is already implicit in the formulation of $\Rc$ and $\Tc$.
If (ii) did not hold, \erf{eq:rijdef} would be ill-defined.  
\medskip

Thus, starting from a defect $D$ obeying (i) and (ii), all defects
generated from $D$ have the same value of $\Rc$. Conversely, one can now ask
whether for a given value $\Rc_0$, one can find a preferred defect
$D_0$ with $\Rc(D_0)=R_0$, such that all elementary conformal defects $D'$ 
with $\Rc(D')=\Rc_0$ can be generated by the
action of topological defects on $D_0$.

For the examples in section \ref{sec:minmod}, where all conformal
defects are known, the answer is yes. In the example treated in
section \ref{sc:Coset}, this is still true for all conformal defects
that preserve a certain extended chiral algebra, but we cannot make
statements about the behaviour of all conformal defects.  For the
Ising model (section \ref{sec:Ising}) it is true for the defects with
discrete excitation spectrum (so that there is an unambiguous notion
of `elementary defect'); it would however not hold true for the
exceptional defects about which we speculate at the end of that section.

\sect{The free boson}\label{sec:freebos}

The simplest model in which conformal defects have been studied is a
single free scalar boson which was investigated in \cite{Bachas:2001vj}. 
In that paper the defect was placed vertically in the plane and the
scalar field to the left and right of the defect were denoted $\phi^1$
and $\phi^2$ respectively and related at the defect by
\be
 \begin{pmatrix} \partial_- \phi^1 \\ \partial_+ \phi^2 
\end{pmatrix}
= S \begin{pmatrix} \partial_+ \phi^1 \\ \partial_- \phi^2 
\end{pmatrix}
\;,\;
\ee
where $S$ is either of the two matrices
\be
 S = \begin{pmatrix}-\cos(2\theta) & \sin(2\theta) \cr
 \sin(2\theta)  & \cos(2\theta)\end{pmatrix}
\;,\;\;
 S' = \begin{pmatrix}\cos(2\theta) & -\sin(2\theta) \cr
 \sin(2\theta)  & \cos(2\theta) \end{pmatrix}
\;.
\label{eq:SSpdef}
\ee
In the folded picture the boundary state representing this defect is 
\begin{equation}
  \begin{split}
 |b\rangle 
  &\;=\; 
  \Nc 
  \prod_{n=1}^\infty
  \exp\Bigl(\;\frac 1n a^i_{-n} \bar a^j_{-n} S_{ij} \Bigr)| 0 \rangle\\
&\;=\; 
  \Nc 
  \left(
  1 
+ a^i_{-1} \bar a^j_{-1} S_{ij} 
+ \frac 12 (a^i_{-1} \bar a^j_{-1} S_{ij})^2
+ a^i_{-2} \bar a^j_{-2} S_{ij} 
+ \ldots
  \right)
  | 0 \rangle
  \end{split}
\end{equation}
where $a^i_n$ and $\bar a^i_n$ are the modes of $\phi^i$ and
$\Nc$ is a normalisation constant. The modes themselves are normalised
such that $[a_m,a_n] = m \delta_{m+n,0}$. Since $c_1=c_2=1$ and the
energy-momentum tensors are of the standard form,  
\be
L^i_{-2}|0\rangle = \frac 12 a^i_{-1}a^i_{-1}|0\rangle
\;,
\ee
it is easy to calculate
\be
 \cev 0 L^i_{2} \ol L{}^j_{2}\vec b =
 \frac {\Nc}2 \,(S_{ij})^2 
\ee
so that for both cases (\ref{eq:SSpdef})
\be
\cev 0 L^{\text{tot}}_{2} \bar L^{\text{tot}}_{2}\vec b
=
\frac{\Nc}2 \sum_{i,j} (S_{ij})^2
= \Nc
\;,
\ee
and
\be
\Rc = \cos^2(2\theta)
\;,\;\;
\Tc = \sin^2(2\theta)
\;.
\ee
  These are the reflection and transmission probabilities for the
  massless modes in this model. Note that the quantity $\Rc$ we have
  defined is {\em not} the same
  quantity as the $\Rc$ calculated in \cite{Bachas:2001vj} which is the
  reflection {\em amplitude} for a massless mode.

\sect{The Ising model and the free fermion}\label{sec:Ising}

The name of Ising model is given to various different theories with
$c=1/2$: the theory of purely local fields of the modular invariant
theory; the local theory of the free fermion; the non-local theory
obtained by combining the two.  In section \ref{ssec:ainv} we discuss
the defects in the modular invariant theory and in section
\ref{ssec:ff} the defects in the free fermion and their relation to
those in the modular invariant theory.  From here on, when we refer
to ``the Ising model'' we shall always mean the modular invariant
theory.

\subsection{The Ising model}\label{ssec:ainv}

Conformal defects in the Ising model have been studied exhaustively
by Oshikawa and Affleck in \cite{Oshikawa:1996ww,Oshikawa:1996dj}.\footnote{
From a lattice or spin chain perspective, the critical properties of  
the Ising model with defect lines have also been studied earlier, see  
e.g.\ \cite{McCoy:1980ag,Henkel:1986dm} and the references in
\cite{Oshikawa:1996ww,Oshikawa:1996dj}. For our analysis we only need
the boundary states in the folded model as first given by Oshikawa
and Affleck.
}
The Ising model
has central charge $c=1/2$ so that the doubled model used in the
folded treatment of conformal defects has central charge one. Oshikawa
and Affleck use two different identifications of this $c=1$ model,
firstly as a special case of the Ashkin-Teller model and secondly as a
particular case of the orbifolded free boson.  We shall use only this
second identification for our calculations.  We review briefly this
construction and then calculate the reflection and transmission
coefficients for the conformal defects they have found. Throughout
this section we use the notation of
\cite{Oshikawa:1996ww,Oshikawa:1996dj}.

The doubled Ising model can be identified with the $r=1$ orbifolded
free boson, that is a free boson which classically takes values in the
line segment $[0,\pi]$.

It is usual to construct the orbifolded free
boson starting from a free boson compactified on a circle.
A complete classification of the conformal boundary conditions
for the free boson on a circle has been proposed by Friedan
\cite{Friedan} with some more details given by
Janik \cite{Janik:2001hb} and
Gaberdiel and Recknagel
\cite{Gaberdiel:2001zq}.
Since the radius 1 is not a rational multiple of the self-dual radius
$r=1/\sqrt 2$, according to \cite{Friedan} the conformal boundary
conditions split into three classes: a circle of Dirichlet boundary
conditions on the free boson, a circle of Neumann boundary conditions,
and a line segment of boundary conditions which break the $U(1)$
symmetry of the free boson.
The construction of the orbifolded free boson boundary states from
those of the free boson on a circle is straightforward for the first
two classes; the third class we will return to later.

The Dirichlet boundary
conditions on the orbifolded free boson $\varphi$ are
$\varphi=\varphi_0\in [0,\pi]$ and the Neumann boundary conditions
can be expressed in terms of the dual field $\tilde\varphi$ as
$\tilde\varphi=\tilde\varphi_0\in[0,\pi/2]$. These boundary conditions
are elementary except at the end points where the presence of twisted
sectors splits them into two. This leads to the following space of
boundary conditions:
\be
  D_O(\varphi_0)\ \text{ with }\ \varphi_0\in(0,\pi)
\;,\;\;
  D_O(0)_\pm
\;,\;\;
  D_O(\pi)_\pm
\ee
\be
  N_O(\tilde\varphi_0)\ \text{ with }\ \tilde\varphi_0\in(0,\pi/2)
\;,\;\;
  N_O(0)_\pm
\;,\;\;
  N_O(\pi/2)_\pm\ \ .
\ee
To construct the space of states we first describe the space
of ground states of 
the un-orbifolded free boson on a circle of radius 1. These are 
labelled by two integers $(m,n)$ and denoted
$\vec{(m,n)}$; the winding number is $m$ and the total momentum is $n$
and the conformal dimensions of such a state are 
$h=(m+\tfrac 12n)^2/2,\bar h=(-m+\tfrac 12n)^2/2$. 
On these ground states the oscillators have integer modes.
In the twisted sector the oscillators have half-integer modes and the
twisted ground states are $\vec{0}_T$ and $\vec{\pi}_T$ with conformal
dimension $1/16$.

The calculation of $\X$ can be reduced to finding the overlap of the
boundary state with a particular bulk state, $\vec{W\ol W}$, as
explained in appendix \ref{app:RT-omega}.  The form of the primary
state $\vec{W\ol W}$ can be deduced from consideration of the
Ashkin-Teller formulation in which it is primary of weight $(2,2)$ with
respect to the total Virasoro algebra and a descendent of the product
of the identity representations of the two $c=1/2$ Virasoro algebras
under the action of the two separate Virasoro algebras.  In
\cite{Oshikawa:1996dj}, such a primary state is denoted by
$\vec{2,II}$ and also by $\vec{2,1}$. {}From a comparison of the
boundary states of the Ising$\times$Ising model constructed in
\cite{Oshikawa:1996dj}, in the Ashkin-Teller and free boson
formulations, the form of the state $\vec{2,II}$ in the orbifolded
free boson construction can be identified, and fixing the
normalisation as in \eqref{eq:W-Wbar-def} we get
\be
 \vec{W\ol W} 
= \frac{1}{8}\,\vec{2,II}
= \frac 1{16}\,
 \Big(\,\vec{(0,4)} + \vec{(0,-4)} + \vec{(2,0)} + \vec{(-2,0)}\,\Big)
\;.
\ee
Calculating $\X$ and 
using the formulae $\Rc {=} (1{+}\X)/2, \Tc{=} (1{-}\X)/2$ we get from
\eqref{eq:RT-via-omega} in this case, 
we find for the various boundary conditions:
\be
\begin{array}{c|ccc}
   & \X & \Rc & \Tc \\
\hline\\[-4mm]
D_O(\varphi_0) 
   & \cos(4\varphi_0) 
   & \cos^2(2\varphi_0)
   & \sin^2(2\varphi_0)
\\ [1mm]
N_O(\tilde\varphi_0) 
   & \cos(4\tilde\varphi_0) 
   & \cos^2(2\tilde\varphi_0)
   & \sin^2(2\tilde\varphi_0)
\\ 
\end{array}
\ee
Note that these also hold true at the end points; the twisted sectors
make no difference to the values of $\X$, $\Rc$ or $\Tc$.

The purely transmitting defects correspond to the Ising model
topological defects as identified in
\cite{Oshikawa:1996ww,Oshikawa:1996dj} -- the trivial or identity
defect is $\Bone = D_O(\pi/4)$, the spin-reversal is $\eps =
D_O(3\pi/4)$ and the duality defect is $\sigma=N_O(\pi/4)$. 

The purely reflecting defects correspond to pairs of boundary
conditions in the two copies of the Ising model. We find a small
difference from the identification given in 
    \cite{Oshikawa:1996dj}, 
in that the Neumann factorising boundary conditions 
have left and right factors swapped. If
we label the three boundary conditions of the Ising model as $+$, $-$ 
and $f$ for spin up, spin down and free, then the identification of
the reflecting defects in the doubled model is: 
\be
\begin{array}{c|c|c|c|c|c|c|c|c}
  (++) & 
  (--) & 
  (-+) & 
  (+-) & 
  (ff) & 
  (f+) & 
  (f-) & 
  (+f) & 
  (-f) \\[2mm] \hline &&&&&&&& \\[-2mm]
  D_O(0)_+ 
& D_O(0)_- 
& D_O(\pi)_+ 
& D_O(\pi)_- 
& D_O(\pi/2) 
& N_O(0)_+ 
& N_O(0)_- 
& N_O(\tfrac{\pi}{2})_+ 
& N_O(\tfrac{\pi}{2})_- 
\end{array}
\nonumber
\ee
This can be checked by computing the overlaps of the boundary states
or by calculating the form of the topological defects in the two
copies of the Ising model and their actions on the boundary states.
Since this seems of some interest, we give the details here.

The topological defects in CFTs $A$ and $B$ on  either side of a  
conformal defect are also topological defects in the product theory
$A\times\bar{B}$ in which the conformal defect is represented by a boundary  
state. We identify these `product defects' in the case of Ising$\times$Ising
by writing down the general ansatz for a topological defect  
and fixing the coefficients by computing the action on the  
factorising boundary states (representing factorising defects).

We do not need to construct the whole topological defect
operator -- only that part which has non-zero action on the boundary
states we are interested in. The boundary states are linear
combinations of Virasoro Ishibashi states, one for each
 spinless primary state (those states which are highest weight
for both the left and right Virasoro algebras with $h=\bar h$) which are
given in \cite{Oshikawa:1996dj}. We repeat the list here but giving
the primary states in the free boson form rather than the
Ashkin-Teller notation in \cite{Oshikawa:1996dj}. For 
$n \in \Zb_{\ge 0}$, 
\be
\begin{array}{c|c}
h=\bar h & \text{Highest weight states}\\ \hline
\\[-3mm]
  n^2 
& 
  \vec n^{\text{Vir}} 
\\[.4em]
  \tfrac 12 (n{+}1)^2
&
  \tfrac{1}{\sqrt 2}\big( \vec{(0,2n{+}2)} + \vec{(0,-2n{-}2)} \big)
\;,\;\;
   \tfrac{1}{\sqrt 2}\big( \vec{(n{+}1,0)} + \vec{(-n{-}1,0)} \big)
\\[.4em]
  \tfrac 18 (2n {+} 1)^2
&
  \tfrac{1}{\sqrt 2}\big( \vec{(0,2n{+}1)} + \vec{(0,-2n{-}1)} \big)
\\[.4em]
  \tfrac 1{16} (2n{+}1)^2
&
  \vec{n}^{\text{Vir}}_{0,T}
\;,\;\;
  \vec{n}^{\text{Vir}}_{\pi,T}
\end{array}
\labl{eq:hhbartable}
The states $\vec{n}^{\text{Vir}}$ are highest weight combinations of
free boson modes on the untwisted vacuum and 
$\vec{n}^{\text{Vir}}_{\varphi_0,T}$ are highest weight combinations of
free boson modes on the twisted vacua, for example
\be
  \vec{1}^{\text{Vir}} = a_{-1} \bar a_{-1} \vac
\;,\;\;
  \vec{1}^{\text{Vir}}_{\pi,T} = a_{-1/2} \bar a_{-1/2} \vec{\pi}_T
\;.
\labl{eq:vacs}
We will write the projectors onto the Virasoro representations with
$h$ equal to $n^2$ and $(2n+1)^2/8$ as $P_n$ and $Q_n$ respectively,
and the matrix part of the topological defect acting on the
representations of weight $(n+1)^2/2$ 
in the basis \erf{eq:hhbartable} as
\be
 \begin{pmatrix} a & b \\ c & d
 \end{pmatrix}_{n}\ \ .
\ee
Finally, the matrix part of the topological defect mixing the twisted
sectors in the basis \erf{eq:hhbartable} we write as
\be
 \begin{pmatrix} a & b \\ c & d
 \end{pmatrix}_{n,T}\ \ .
\ee
Thus, we consider defects of the form
\be
\hat X=
\sum_{n=0}^\infty\left(
  a_n P_n
+ b_n Q_n
+ \begin{pmatrix} c_n & d_n \\ e_n & f_n
  \end{pmatrix}_{n}
+ \begin{pmatrix} g_n & h_n \\ j_n & k_n
  \end{pmatrix}_{n,T}\right)
+ \hbox{\it other terms}
\ee
where `{\it other terms}' are parts which annihilate the conformal
boundary states.

One can deduce from \cite{Oshikawa:1996dj} or from the character
formulae for the vacuum and twisted free boson  
representations that the free boson Dirichlet and
Neumann Ishibashi states are given in terms of the Virasoro Ishibashi
states as follows:
\be
\begin{array}{rclrcl}\displaystyle
      |\,(0,0)\,\rrangle_D
\etb = \etb \sum_{n=0}^\infty \vecc{n}^\tV \;,
\quad\etb
      \vecc{\phi}_{T,D}
\etb = \etb \sum_{n=0}^\infty \vecc{n}^\tV_{\phi,T} \;,
\enl
      |\,(0,0)\,\rrangle_N 
\etb = \etb \sum_{n=0}^\infty (-1)^n \vecc{n}^\tV \;,
\quad\etb
      \vecc{\phi}_{T,N}
\etb = \etb \sum_{n=0}^\infty (-1)^{n(n+1)/2}
\vecc{n}^\tV_{\phi,T} 
\;.
\end{array}
\labl{eq:Ishibashis}
where $\phi\,{\in}\{0,\pi\}$ labels the twisted sector.
The identification of the factorised boundary conditions in the
Dirichlet sector fixes the forms of the identity and $\eps$
defects as
\be
\begin{array}{rl}\displaystyle
 \Bone_L = \Bone_R =
&\displaystyle
\sum_{n=0}^\infty
\left(
 P_n + Q_n + \begin{pmatrix} 1 & 0 \\ 0 & 1 \end{pmatrix}_{n}
           + \begin{pmatrix} 1 & 0 \\ 0 & 1 \end{pmatrix}_{n,T}
\right)
+ \ldots
\\[5mm]
 \eps_L = 
&\displaystyle
 \sum_{n=0}^\infty
  \left(
  P_n - Q_n +  \begin{pmatrix} 1 & 0 \\ 0 & 1 \end{pmatrix}_{n}
            +  \begin{pmatrix} 0 & 1 \\ 1 & 0 \end{pmatrix}_{n,T}
\right)
+ \ldots
\\[5mm]
 \eps_R = 
&\displaystyle
 \sum_{n=0}^\infty
  \left(
  P_n - Q_n +  \begin{pmatrix} 1 & 0 \\ 0 & 1 \end{pmatrix}_{n}
            +  \begin{pmatrix} 0 & -1 \\ -1 & 0 \end{pmatrix}_{n,T}
\right)
+ \ldots
\eear
\ee
These satisfy $\eps_L^2 = \eps_R^2 = 1$, and the total spin
reversal, given as the product of the reversals in the two sectors,
\be
  \eps_L\eps_R
= 
\sum_{n=0}^\infty \left(
  P_n + Q_n +  \begin{pmatrix} 1 & 0 \\ 0 & 1 \end{pmatrix}_{n}
            +  \begin{pmatrix} -1 & 0 \\ 0 & -1 \end{pmatrix}_{n,T}
\right)
+ \ldots
\;,
\ee
does just reverse the sign of the twisted sector contributions as
noted in \cite{Oshikawa:1996ww,Oshikawa:1996dj}.

To complete the set of topological defects, we find the duality
defects are
\be
 \begin{array}{rl}\displaystyle
\sigma_L = 
&\displaystyle 
\sqrt2 \sum_{n=0}^\infty \left(
  (-1)^n P_n + \begin{pmatrix} 0 & 1 \\ 1 & 0 \end{pmatrix}_{n}
             + \frac12 {(-1)^{n(n+1)/2}}
               \begin{pmatrix} 1 & 1 \\ 1 & 1 \end{pmatrix}_{n,T}
\right) + \ldots
\\[5mm]
\sigma_R = 
&\displaystyle 
\sqrt2 \sum_{n=0}^\infty \left(
  (-1)^n P_n + (-1)^n\begin{pmatrix} 0 & 1 \\ 1 & 0 \end{pmatrix}_{n}
             + \frac 12 {(-1)^{n(n+1)/2}}
               \begin{pmatrix} 1 & -1 \\ -1 & 1 \end{pmatrix}_{n,T}
\right) + \ldots
\;.
\eear
\ee
Again it is easy to check that these obey the correct algebra:
\be
 \sigma_L^2 = \Bone + \eps_L
\;,\;\;
 \sigma_R^2 = \Bone + \eps_R
\;,\;\;
 \sigma_L \eps_L = \sigma_L
\;,\;\;
 \sigma_R \eps_R = \sigma_R
\;.
\ee

We can now investigate the action of the topological defects on
the 
    Dirichlet and Neumann line
of boundary conditions in the orbifolded model:
\be
\begin{array}{ll}\displaystyle
  \eps_L \vec{D_O(\varphi)}
= \eps_R \vec{D_O(\varphi)}
= \vec{ D_O(\pi-\varphi)}
\;,\etb
  \eps_L \vec{N_O(\tilde\varphi)}
= \eps_R \vec{N_O(\tilde\varphi)}
= \vec{ N_O(\tilde\varphi)}
\;, 
\\[.4em]\displaystyle
  \sigma_L \vec{D_O(\varphi)}
= \sigma_L \vec{D_O(\pi{-}\varphi)}
= \vec{ N_O(\varphi)} 
\;,\etb
  \sigma_R \vec{D_O(\varphi)}
= \sigma_R \vec{D_O(\pi{-}\varphi)}
= \vec{ N_O(\tfrac{\pi}{2}{-}\varphi)} 
\;,
\\[.4em]\displaystyle
  \sigma_L \vec{N_O(\tilde\varphi)}
= \vec{ D_O(\tilde\varphi)} 
+ \vec{ D_O(\pi - \tilde\varphi)}
\;,\etb
  \sigma_R \vec{N_O(\tilde\varphi)}
= \vec{ D_O(\tfrac{\pi}{2}-\tilde\varphi)} 
+ \vec{ D_O(\tfrac{\pi}{2}+ \tilde\varphi)}
\;.
\eear
\ee
We can write this in terms of defect fusion as 
\be
\begin{array}{ll}\displaystyle
 X_\eps \star D_O(\varphi) = 
 D_O(\varphi) \star X_\eps =
 D_O(\pi - \varphi)
\;,\etb
 X_\eps \star N_O(\tilde\varphi) = 
 N_O(\tilde\varphi) \star X_\eps =
 N_O(\tilde\varphi)
\;,
\\[.4em]\displaystyle
 X_\sigma \star D_O(\varphi) = 
 N_O(\varphi)
\;,\etb
 D_O(\varphi)\star X_\sigma = 
 N_O(\tfrac{\pi}{2}-\varphi)
\;,
\\[.4em]\displaystyle
 X_\sigma \star N_O(\tilde\varphi) = 
 D_O(\tilde\varphi) + D_O(\pi-\tilde\varphi)
\;,\etb
 N_O(\tilde\varphi) \star X_\sigma  = 
 D_O(\tfrac{\pi}{2}-\tilde\varphi) + D_O(\tfrac{\pi}{2}+\tilde\varphi)
\;.
\eear
\ee
This means that the number $\X$ does not uniquely specify the orbit
under the action of the group-like topological defects, nor do the
pair $(\X,g)$ where $g$ is the ground-state-degeneracy of the boundary
condition in the folded model as indicated in figure \ref{fig:TwoSines}.
For a given pair $(\X,g)$, there are two
orbits of the group-like defects.
On the other hand, starting from any one elementary defect 
            on the Dirichlet or Neumann line
at a given
value of $\X$, all other elementary defects with this $\X$
can be generated by the
action of topological defects (in the sense explained in section 
\ref{sec:RT-top}).

In figure \ref{fig:TwoSines} we give a pictorial summary of the value
of $\X$ and the actions of $\eps$ and $\sigma$ on the Dirichlet and
Neumann line of conformal defects in the Ising model. 

\FIGURE[t]{
  \begin{picture}(300,280)
    \put(-36,1){ 
    \scalebox{1}{\includegraphics{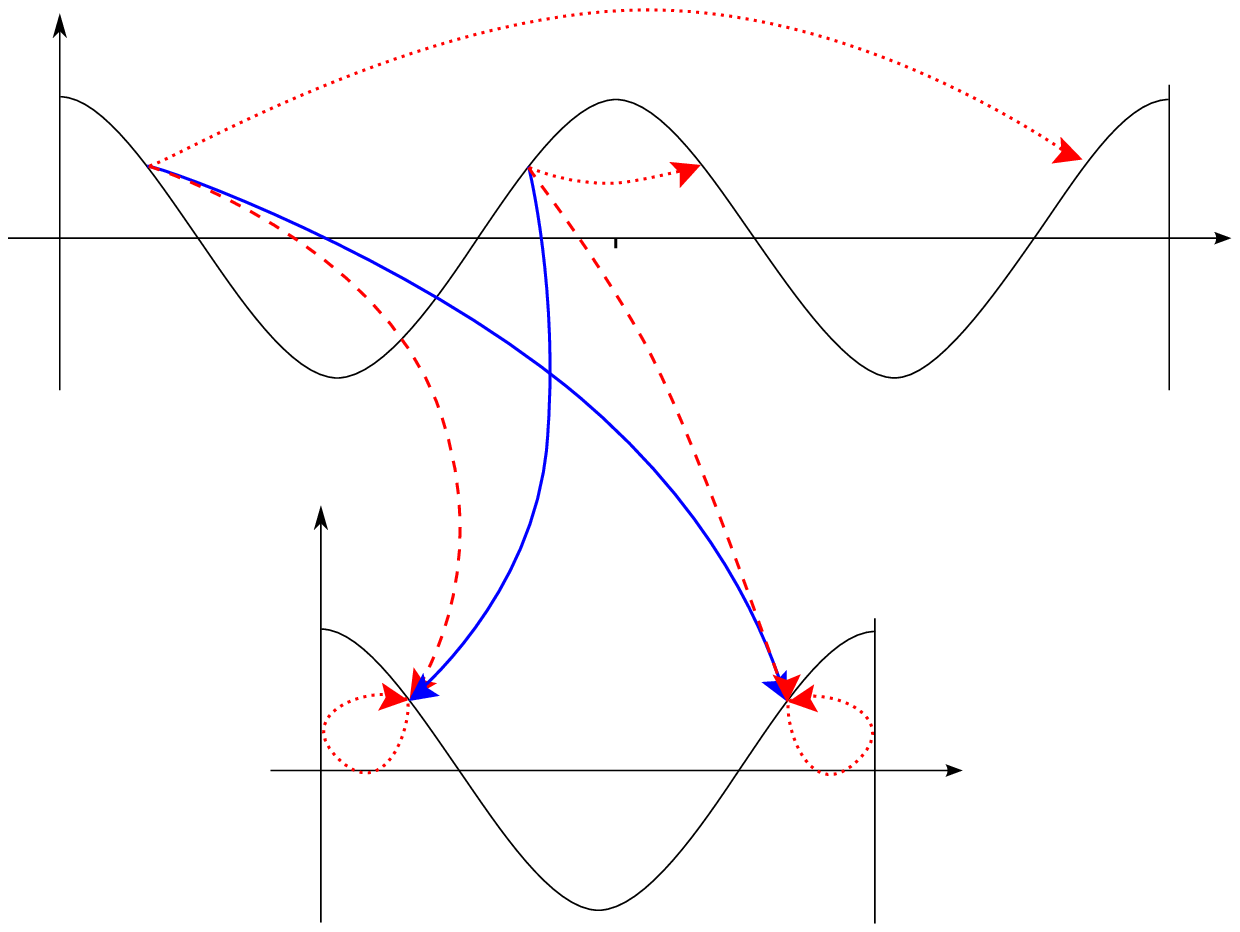}} }
    \setlength{\unitlength}{1pt}
    \put( 5,84)   {\scriptsize$(f\,+)\cup(f\,-)$}
    \put( 225,84)   {\scriptsize$(+\,f)\cup(-\,f)$}
    \put( -75,239)   {\scriptsize$(+\,+)\cup(-\,-)$}
    \put( 310,239)   {\scriptsize$(+\,-)\cup(-\,+)$}
    \put( 136,245)   {\scriptsize$(f\,f)$}
    \put( 60,150)   {\scriptsize$\Bone$}
    \put(222,150)   {$\eps$}
    \put(139,-3)   {$\sigma$}
    \put(-20,270)   {$\X$}
    \put(55,125)   {$\X$}
    \put(327,198)   {$\varphi$}
    \put(251,45)   {$\tilde\varphi$}
    \put(140,204)   {\scriptsize$\pi/2$}
    \put(307,204)   {\scriptsize$\pi$}
    \put(223,50)   {\scriptsize$\pi/2$}
    \put( -70,200)   {$g=1$}
    \put( -70,45)   {$g=\sqrt 2$}
    \setlength{\unitlength}{1pt}
  \end{picture}
\caption{\label{fig:TwoSines}The values of $\X$ and $g$, and the action of
the topological defects on the Dirichlet and Neumann lines of
conformal defects in the Ising model.  The action of $X_\eps$ from
the left and from the right is the same and is given by the dotted
arrows, the action of $X_\sigma$ on the left by the dashed arrows and
on the right by the solid arrows. Note that the end points of the
Dirichlet and Neumann lines are superpositions of two elementary
defects as indicated.}  
}

Finally we return to the question of extra defects in the orbifold
model. Neither the classification of boundary states of \cite{Friedan}
nor the explicit calculations of \cite{Gaberdiel:2001zq,Janik:2001hb} 
appear to have been
performed for the orbifold model, so the comments here remain
speculative. The line of exceptional boundary states described in
\cite{Friedan,Gaberdiel:2001zq,Janik:2001hb} 
are invariant under the orbifold action
$\varphi\mapsto-\varphi$ so we might expect that they would lead to
two sets of defects by the addition of twisted sectors, but since both
the exceptional boundary states and all states in the twisted sectors
have zero overlap with $\vec{W\bar W}$ we would expect them to have $\X=0$
and hence $\Rc{=}\Tc{=}1/2$.

\subsection{Free fermions}\label{ssec:ff}

As stated in the beginning of this section, by the Ising model we
denote the modular invariant theory. Free fermions can then be
understood as non-local or disorder fields in the Ising model. Defects
for free fermions have been considered before, for example in
\cite{Delfino:1994nr}. 

The simplest class of conformal defects in the free fermion model are
those that satisfy
\be
\bearrl
  \psi_m\, \hat D
&=\; 
   T^{11}\, \hat D\, \psi_m
 \;+\; 
   T^{12}\, \bar\psi_{-m}\, \hat D 
\;,\;\;
\enl
  \hat D\, \bar\psi_{-m}
&=\; 
    T^{21}\, \hat D\, \psi_m  
  \;+\; 
    T^{22}\, \bar\psi_{-m}\, \hat D
\;.
\eear
\labl{eq:ffdefect}
Compatibility with the fermion algebra implies that the matrix $T$
takes the form
\be
  T = \begin{pmatrix}
  \cos(\chi) & i \sin(\chi) \\
  i\eta\sin(\chi) & \eta\cos(\chi)
 \end{pmatrix}
\labl{eq:TRdef}
where $\eta=\pm 1$.
Reference \cite{Delfino:1994nr} deals primarily with massive fermions
in the defect 
picture but it is easy to take the massless limit and they find that
the defects 
       treated there
all satisfy \eqref{eq:ffdefect} and \eqref{eq:TRdef} with
$\eta=1$.

The defects in the local Ising model are not expected to be in one-to-one
correspondence with those in the free fermion model, but they are
related as the Ising model contains certain combinations of the free
fermion fields, notably the primary field
\be
 \eps = i \bar\psi\psi\;\;.
\labl{eq:epsdef}
This enables one to determine the following combinations:
\bea
 T^{12} 
= \cev 0 \bar\psi_{1/2}\psi_{1/2} \hat D \vac / \cav \hat D \vac
= -i\cev{\eps}\hat D \vac / \cav \hat D \vac
\enl
 T^{21} 
= \cav \hat D \bar\psi_{-1/2}\psi_{-1/2} \vac / \cav \hat D \vac 
= -i\cav \hat D \vec{\eps} / \cav \hat D \vac
\enl
 T^{12}T^{21}-T^{11}T^{22}
= \cav  \bar\psi_{1/2}\psi_{1/2} \hat D \bar\psi_{-1/2}\psi_{-1/2} \vac  
  / \cav  \hat D \vac
= -\cev{\eps} \hat D \vec{\eps} / \cav \hat D \vac\;\;.
\eear
\labl{eq:ssfound}
To calculate these using the boundary state formalism we need to
identify the fields $\eps^1,\eps^2$ in the doubled model corresponding
to the field $\eps$ on the left and on the right of the defect.
The fields $\eps^i$ both have conformal dimensions $(\tfrac 12,\tfrac
12)$ and there are two primary fields with these weights in the
orbifolded free boson. If we use the convention of
\cite{Delfino:1994nr} that the reflection factor for 
fermion modes at a `fixed' boundary is $-i$ and at a `free' boundary
is $i$ then we find agreement with our assignment of factorising
boundary conditions with the choice 
\bea
\vec{\eps^1} = \tfrac 12 \Big(
  \vec{(0,2)} + \vec{(0,-2)} - \vec{(1,0)} - \vec{(-1,0)}\Big)
\;,\;\;\\[3mm]
\vec{\eps^2} = \tfrac 12 \Big(
  \vec{(0,2)} + \vec{(0,-2)} + \vec{(1,0)} + \vec{(-1,0)}\Big)
\;,\;\;\\[3mm]
\vec{\eps^1\eps^2} = a_{-1}\bar a_{-1}\vac
\eear
\labl{eq:epsdefs}
where $\vec{(m,n)}$ are the free boson vacuum states described
earlier and $a_{-1}$ and $\bar a_{-1}$ are modes of the free boson
field.

With these choices, one finds for the
Neumann defects $N_O(\tilde\varphi_0)$ that
\be
T^{11}=-T^{22}=i\cos(2\tilde\varphi_0)
\;,
\labl{eq:s1122n}
so that these cannot be identified with any of the defects in
\cite{Delfino:1994nr} for 
any value of $\chi$ since they all have $T^{11}=T^{22}$.

For the Dirichlet defects $D_O(\varphi_0)$, however,
\be
T^{12}=T^{21}=-i\cos(2\varphi_0)
\;,\;\;
T^{11}T^{22} = \sin^2(2\varphi_0)
\;,
\labl{eq:s1122d}
      so that $\eta=1$ in \erf{eq:TRdef}, and hence also
      $T^{11}=T^{22}$. These 
      defects can thus
be identified with the defects in \cite{Delfino:1994nr} with
\be
\varphi_0 = \pm \tfrac 12(\chi + \pi/2)
\;.
\ee
The ambiguity in $\chi$ corresponds to the ambiguity in the
definition of the fermion fields.
As a final comment we note that our quantities $\Rc$ and $\Tc$ can
again be identified as the reflection and transmission {\em
probabilities} for the modes of the free fermion.

\sect{Conformal defects from coset decompositions}\label{sec:coset}

Consider two conformal field theories $\text{CFT}_1$ and $\text{CFT}_2$ 
which are joined along a defect, and denote their chiral algebras 
by $\mc{A}_1$ and $\mc{A}_2$, respectively. For
  each embedding of another chiral algebra $\mc{B}$ into the symmetry
  algebra $\mc{A}_1\otimes\mc{A}_2$ of the product theory
  $\text{CFT}_1\otimes\overline{\text{CFT}}_2$ one can
  then construct a family of non-trivial conformal defects in the
  following way. First, we analyse the original theory with respect to
  the reduced symmetry
\begin{equation}
  \label{eq:Deco1}
  \mc{A}_1\otimes\mc{A}_2\ \to\ 
  \mc{B}\otimes(\mc{A}_1\otimes\mc{A}_2)/\mc{B}\ \ ,
\end{equation}
  where the coset chiral algebra $(\mc{A}_1\otimes\mc{A}_2)/\mc{B}$
  contains the fields which commute with all the operators in
  $\mc{B}$.\footnote{
  The interpretation of this decomposition is most straightforward
  if the theories on the two sides of the defect are WZW models based
  on some groups $G_1$ and $G_2$. In that case one can define the
  chiral subalgebra $\mc{B}$ in terms of a WZW model related to a common
  subgroup of $G_1$ and $G_2$. The coset part on the other hand is
  described by the standard GKO construction.}
  On the practical level reducing the symmetry means that we have to
  decompose the Hilbert space of the folded model in terms of
  representations of the smaller symmetry. The theory itself obviously
  does not change. 
\smallskip

  In a second step we impose boundary conditions which just preserve
  the chiral algebra on the right hand side of \eqref{eq:Deco1}.
  This procedure defines defects between $\text{CFT}_1$ and
  $\text{CFT}_2$, 
  which in general are neither fully reflective nor fully transmissive
  \cite{Quella:2002ct}.
  Note that the interpretation of which excitations are transmitted and
  which reflected is somewhat mysterious since the coupling of the two
  theories takes place in the coset part. There is, however, one special
  class of defects where the physical interpretation is immediately
  obvious and a concrete formula for the transmission can easily be
  derived. This will be the subject of the following subsection. Later
  we will return to the general case and discuss one example in
  detail.

\subsection{Transmission of a common sub-symmetry}\label{sec:SubSym}

  One of the most intuitive examples of semi-permeable defect lines
  arises if the conformal field theories on the two sides of the defect
  share a common chiral subalgebra $\mc{C}$. In that case we can employ
  the previous construction with $\mc{B}=\mc{C}\otimes\mc{C}$ such that
  the first factor $\mc{C}$ is embedded into $\mc{A}_1$ and the second
  into $\mc{A}_2$. It is then straightforward to construct conformal
  defects which allow all excitations 
  with respect to the sub-symmetry
  $\mc{C}$ to pass through unaffectedly. In the folded picture this
  corresponds to boundary conditions which just preserve the sub-symmetry
\begin{equation}
  \label{eq:Deco2}
  \mc{A}_1\otimes\mc{A}_2\ \to\ 
  \mc{A}_1/\mc{C}\otimes\mc{A}_2/\mc{C}\otimes\mc{C}\otimes\mc{C}
\end{equation}
  of the full bulk system and where one has permutation type boundary
  conditions in the product of the two $\mc{C}$-sectors.
  In the two coset sectors on the other hand
  we assume trivial gluing conditions.
\smallskip

  In order to determine the transmission we need to calculate the
  correlation functions between the fields 
  $T^1=T^{\mc{A}_1/\mc{C}}+T^{1,\mc{C}}$ and
  $T^2=T^{\mc{A}_2/\mc{C}}+T^{2,\mc{C}}$ as well as their
  anti-holomorphic counterparts in the presence of the boundary.
  On the level of the individual energy momentum tensors our
  choice of
          boundary condition  
  corresponds to the gluing conditions
\begin{align}
  T^{\mc{A}_1/\mc{C}}&\ =\ \ol T{}^{\mc{A}_1/\mc{C}}&
  T^{\mc{A}_2/\mc{C}}&\ =\ \ol T{}^{\mc{A}_2/\mc{C}}&
  T^{1,\mc{C}}&\ =\ \ol T{}^{2,\mc{C}} &
  T^{2,\mc{C}}&\ =\ \ol T{}^{1,\mc{C}}
\end{align}
  on the real axis. These allow us to
  simplify the boundary correlation functions considerably.
  For instance we obtain
\be\bearll
  \cev 0 L^1_2 \ol L{}^2_2 \vec b 
  \etb =\ \cev 0 \bigl(L^{\mc{A}_1/\mc{C}}_2 + L^{1,\mc{C}}_2\bigr)
  \bigl(\ol L{}^{\mc{A}_2/\mc{C}}_2 + \ol L{}^{2,\mc{C}}_2\bigr)\vec b
\\[.5em]
  \etb =\ \cev 0 \bigl(L^{\mc{A}_1/\mc{C}}_2+L^{1,\mc{C}}_2\bigr)
  \bigl(L^{\mc{A}_2/\mc{C}}_{-2} + L^{1,\mc{C}}_{-2}\bigr)\vec b
  \ =\ \tfrac12 \, c \, \langle 0| b \rangle \ .
\eear\ee
  Here, $c$ denotes the Virasoro central charges of $\mc{C}$ and
  we will similarly use the symbols $c_{1,2}$ for the central
  charges of the chiral algebras $\mc{A}_{1,2}$. It is then a simple
  exercise to plug these expressions into the defining formula
  \erf{eq:RT-via-omega} for the transmission, yielding
\begin{equation}
  \label{eq:SubTrans}
  \Tc\ =\ \frac{2c}{c_1+c_2}
\end{equation}
  for this specific type of defect. In accordance with physical
  intuition the quantity \eqref{eq:SubTrans} takes values between $0$
  and $1$, at least for unitary theories. Moreover, it is maximised 
  if the chiral algebra of one of the models is contained in the one of
  the other. We would like to stress that in our derivation no
  reference has been made to the concrete form of the boundary states
  or similar data. The reader who is interested in these objects
  should consult the references \cite{Quella:2002ct,Quella:2002ns} for
  further details of the construction.
\smallskip

In order to gain some intuition let us consider a defect between
two WZW models based on the group $SU(2)$ and with levels $k_1$
and $k_2$, respectively. Both WZW models share a common sub-symmetry
$U(1)$.\footnote{In order to conform to the condition that the chiral
  subalgebras agree we have to choose the {\em unextended} Heisenberg
  algebra $\mc{C}=\hat{u}(1)$ as the common symmetry.} With the concrete
expressions for the central charges we immediately find
\begin{equation}
  \Tc(k_1,k_2)\ =\ \frac{(k_1+2)(k_2+2)}{3(k_1k_2+k_1+k_2)}\ \ .
\end{equation}
  It may be checked that the transmission takes values in the
  interval between $1/3$ and $1$. Full transmission $\Tc=1$ is
  only achieved when $k_1=k_2=1$. This is not surprising since in this
  case one has the equivalence $SU(2)_1=U(1)_1$, rendering the
  cosets trivial. The other limiting case $\Tc=1/3$ is obtained
  in the regime of large levels. Let us finally state the result
  $\Tc=(k+2)/3k$ which arises if both levels are chosen equal
  to $k$. A diagram of the transmission coefficients
  can be found in figure \ref{fig:Trans}\,a) below.
\smallskip

  Defects of the type considered in this subsection also exist
  if the WZW models on the two sides are based on different groups,
  or for cosets. 
  We list some examples in table~\ref{tb:Transmissions}.

\TABLE[tb]{
\begin{tabular}{cccc}
  $\text{CFT}_1$ 
& $\text{CFT}_2$ 
& Symmetry preserved 
& Transmission $\Tc$
\\\hline\\[-2mm]
  {\scriptsize $(\mc{G}_1)_{k_1}$} 
& {\scriptsize $(\mc{G}_2)_{k_2}$} 
& $\frac{\mc{G}_1\times\mc{G}_2}{U(1)^l\times U(1)^l}$
    {\scriptsize $\times U(1)^l\times U(1)^l$} 
& $\frac{2l(k_1+g_1^\vee)(k_2+g_2^\vee)}
    {k_1(k_2+g_2^\vee)\dim\mc{G}_1+k_2(k_1+g_1^\vee)\dim\mc{G}_2}$ 
\\[2mm]
  {\scriptsize $SU(3)_k$} 
& {\scriptsize $SU(3)_{4k}$} 
& $\frac{SU(3)_k}{SU(2)_{4k}}${\scriptsize$\times$}$
  \frac{SU(3)_{4k}}{SU(2)_{4k}}$
  {\scriptsize $\times SU(2)_{4k}\times SU(2)_{4k}$} 
  \hspace*{-2em} 
& $\frac{3(k+3)(4k+3)}{2(2k+1)(8k+15)}$
\\[2mm]
  {\scriptsize $(F_4)_k$} 
& {\scriptsize $(E_6)_k$} 
& $\frac{(F_4)_k}{(G_2)_k}$
  {\scriptsize$\times$}$\frac{(E_6)_k}{(G_2)_k}$
  {\scriptsize $\times(G_2)_k\times(G_2)_k$} 
& $\frac{14(k+9)(k+12)}{13(k+4)(5k+51)}$ 
\\[2mm]
  {\scriptsize $SU(2)_k$} 
& $\frac{SU(2)_k}{U(1)_k}$ 
& $\frac{SU(2)}{U(1)}${\scriptsize$\times$}
  $\frac{SU(2)}{U(1)}${\scriptsize $\times U(1)$} 
& $\frac{4(k-1)}{5k-2}$
\\[2mm]
  {\scriptsize $SU(2)_{k_1}$} 
& {\scriptsize $SU(2)_{k_2}$} 
& $\frac{SU(2)_{k_1}\times SU(2)_{k_2}}{SU(2)_{k_1+k_2}}$
  {\scriptsize $\times SU(2)_{k_1+k_2}$} 
& \hspace*{-1.5em} 
  $\frac{4k_1k_2(k_1+2)(k_2+2)}{(k_1k_2+k_1+k_2)(k_1+k_2)(k_1+k_2+4)}
  \,\frac{1+\cos\frac{2\pi}{k_1+k_2+2}}{1+2\cos\frac{2\pi}{k_1+k_2+2}}$ 
\end{tabular}
\caption{List of some defect systems and their respective
  transmissions. The first four examples are of the form treated in
  section \ref{sec:SubSym}. The
parameter $l$ in the first line obeys 
$1\leq l\leq\min(\rank\,\mc{G}_1,\rank\,\mc{G}_2)$. In the last
line (see section \ref{sc:Coset})
$k_1$, $k_2$ are odd, and only the maximal value for $\Tc$ is listed.}
\label{tb:Transmissions}
}

\subsection{An example for the general case}\label{sc:Coset}

After the exhaustive discussion of the special case
$\mc{B}=\mc{C}\otimes\mc{C}$ we would like to return to a general
defect involving a symmetry breaking of the form \eqref{eq:Deco1}.
If the subalgebra $\mc{B}$ is not of the above factorised from, and
hence the embedding  of $\mc{B}$ into $\mc{A}_1\otimes\mc{A}_2$ is not
a product of two separate embeddings as in the previous section, 
a physical interpretation regarding which degrees of freedom are
transmitted and which reflected is not obvious. Having access
to the value of the transmission is then a good hint about the actual
nature of the defect.
\smallskip

The calculation of the transmission for decompositions of the
form \eqref{eq:Deco1} requires rather detailed knowledge
about the representation theory of the chiral algebras involved.
For simplicity we will thus specifically consider the coset decompositions
\begin{equation}
  \label{eq:Deco}
  SU(2)_{k_1}\times SU(2)_{k_2}\ \to\ 
  \frac{SU(2)_{k_1}\times SU(2)_{k_2}}{SU(2)_{k_1+k_2}}
  \times SU(2)_{k_1+k_2}
\end{equation}
  in order to explain the main steps. These examples also have
  the advantage of making contact to the previous subsection where
  a different type of defect between the same conformal field theories
  has been discussed. But in contrast to the last case there is no
  general and simple derivation of the transmission amplitude
  available in this case.
  Instead, we are forced into a relatively voluminous and
  model specific calculation starting from the definition
  \eqref{eq:rijdef}.
  The relevant boundary states $|B\rangle$ are given in terms of
  Ishibashi states 
  which preserve the right hand side of \eqref{eq:Deco}
  and which are themselves sums over tensor products of a set of
  orthonormalised states in irreducible representations of the reduced
  symmetry. Consequently, we need to express the states
\begin{equation}
  \label{eq:IntStates}
  L_{-2}^{i} \ol L{}_{-2}^{j}|0\rangle
\end{equation}
  in terms of an orthonormal
  basis of the corresponding representation spaces.
\smallskip

  The states we are interested in reside in the vacuum module
  of $SU(2)_{k_1}\times SU(2)_{k_2}$. In order to identify these
  states after the coset decomposition \eqref{eq:Deco} we have to
  work out the branching functions of the affine representations
  involved. Since the states \eqref{eq:IntStates} have
  $h=2$ and are all singlets with respect to the zero mode algebra
  of $SU(2)_{k_{12}}$, where $k_{12}=k_1+k_2$, we can restrict our
  attention   to the first few characters $\chi_0^{(k_{12})}$,
  $\chi_1^{(k_{12})}$   and $\chi_2^{(k_{12})}$ of the diagonal affine
  subalgebra $SU(2)_{k_{12}}$. 
  When writing down the characters it
  is useful to keep track of the representation content with respect
  to   the $su(2)$ zero mode algebra on each energy level. For
  $k\geq2$ the   relevant characters of $SU(2)_k$ can then easily be
  found to be 
  (up to higher orders in $q$ and not spelling out the overall
  prefactor $q^{h-c/24}$)
\begin{equation}
  \begin{split}
 \chi_0^{(k)}(q)&\ =\ (0)+q\,(1)+q^2\,\bigl((0)+(1)+(2)\bigr)+\cdots
\\[2mm]
 \chi_1^{(k)}(q)&\ =\ (1)+q\,\bigl((0)+(1)+(2)\bigr)+\cdots
\\[2mm]
 \chi_2^{(k)}(q)&\ =\ (2)+q\,\bigl((1)+(2)+(3)\bigr)+\cdots\ \ .
  \end{split}
  \label{eq:su2-char-horiz}
\end{equation}
  Here, the symbols $(j)$ denote an $su(2)$ representation of spin $j$.
  For a first hint towards where to search for the relevant states
  \eqref{eq:IntStates}
  in the decomposed theory it is useful to start with the following
  equality, 
    which involves the affine characters of $SU(2)$ and is valid for
  $k_1,k_2\geq2$,
\begin{equation}
  \begin{split}
    \chi_0^{(k_1)}(q)\,\chi_0^{(k_2)}(q)
    &\ =\ (0)+2\,q\,(1)+3\,q^2\bigl((0)+(1)+(2)\bigr)+\cdots\\[2mm]
    &\ =\ \chi_0^{(k_{12})}(q)\,\bigl(1+q^2+\cdots\bigr)
          +\chi_1^{(k_{12})}(q)\,q\bigl(1+q+\cdots\bigr)\\[2mm]
    &\qquad+\chi_2^{(k_{12})}(q)\,q^2\bigl(1+\cdots\bigr)+\cdots\ \ ,
  \end{split}
  \label{eq:su2-char-prod}
\end{equation}
       where in the first line we decomposed the tensor product of the
       two vacuum modules with respect to the zero mode action of the
       diagonal $SU(2)_{k_{12}}$.
{}From \erf{eq:su2-char-horiz} we see that the representations $(0)$
and $(1)$ of $SU(2)_{k_{12}}$ contain singlets at levels 0 and 2, and
1, respectively, while $(2)$ does not have a singlet at level 0. The
three level 2 singlets in \erf{eq:su2-char-prod} thus lie in
$(0,0,0)\times(0)$ and $(0,0,1)\times(1)$. 
In this notation, the triple refers to
a representation of the diagonal $SU(2)$ coset while the last
label specifies a representation of $SU(2)_{k_{12}}$.
In particular, the desired states \eqref{eq:IntStates}
are a linear combination of maximally three states inside of
$(0,0,0)\times(0)$ and $(0,0,1)\times(1)$.
\smallskip

With a bit of work one finds that the primary states corresponding to 
the representations just mentioned are given by
$|(0,0,0)\rangle\otimes|0\rangle=|0\rangle\otimes|0\rangle$ as well
as\footnote{Our current algebra conventions are summarised in
  eq.~\eqref{eq:CurAlg}.}
\begin{equation}
  \label{eq:SpecState}
  |(0,0,1)\rangle\otimes|1\rangle
  \ =\ \frac{1}{\sqrt{k_1k_2(k_1+k_2)}}\Bigl(k_2E_{-1}^+|0\rangle\otimes|0\rangle
       -k_1|0\rangle\otimes E_{-1}^+|0\rangle\Bigr)\ \ .
\end{equation}
Here, the left hand side refers to the coset decomposition while the right
hand side is the corresponding expression in the product of
affine representations of $SU(2)_{k_1}$ and $SU(2)_{k_2}$.
By acting with creation operators on the two
states above one reveals the orthonormal singlet states $|v_i\rangle$,
$i=1,2,3$, we are interested in.
The first two of these lie in $(0,0,0)\times(0)$ and
correspond via the state field correspondence to the
energy momentum tensors of the coset theory and the affine subalgebra,
respectively. The third state belongs
to a descendent of the affine highest weight state $|1\rangle$. Explicit
formulas for the vectors $v_i$ can be found in the appendix in
eq.~\eqref{eq:CandVectors}.
\smallskip

Using the vectors $v_i$, the states \eqref{eq:IntStates} corresponding to
the energy momentum tensors of the individual theories $SU(2)_{k_1}$ and
$SU(2)_{k_2}$ can be expressed in an adapted basis, see
eqs.~\eqref{eq:States} and~\eqref{eq:StatesSquare}. In order to
calculate the transmission we 
then have to evaluate all the overlaps
\begin{equation}
  \label{eq:Overlap}
  \langle0|L_{2}^{i}\bar{L}_{2}^{j}|b\rangle ~.
\end{equation}
  For the present choice of
  gluing conditions the boundary states are labelled by
  a triple $(\rho_1,\rho_2,\rho)$ whose entries label
  representations of the three affine algebras involved
  \cite{Quella:2002ct},
  i.e.\ they satisfy $2\rho_i\in\{0,1,\cdots,k_i\}$ and
  $2\rho\in\{0,1,\cdots,k_1+k_2\}$. In addition there is an
  identification
  $(\rho_1,\rho_2,\rho)\sim
  \bigl(k_1/2-\rho_1,k_2/2-\rho_2,(k_1+k_2)/2-\rho\bigr)$
  (but no selection rule).
\smallskip

  For the actual calculation of the overlap \eqref{eq:Overlap}
  we need to expand the boundary state in terms of Ishibashi
  states which implement the proper gluing conditions.
  The corresponding expressions are \cite{Quella:2002ct}
\begin{equation}
  \label{eq:BS}
  |(\rho_1,\rho_2,\rho)\rangle
  \ =\ \sum_{\substack{\mu_1,\mu_2,\mu
\\
  \mu_1+\mu_2+\mu\in\Zb}}\,
  \frac{S_{\rho_1\mu_1}^{(k_1)}S_{\rho_2\mu_2}^{(k_2)}}
  {\sqrt{S_{0\mu_1}^{(k_1)}S_{0\mu_2}^{(k_2)}}}
\,\frac{S_{\rho\mu}^{(k_1+k_2)}}{S_{0\mu}^{(k_1+k_2)}}\,
  |(\mu_1,\mu_2,\mu)\rrangle{\otimes}|\mu\rrangle\ \ .
\end{equation}
  The sum runs over the ranges $2\mu_i\in\{0,1,\cdots,k_i\}$ and
  $2\mu\in\{0,1,\cdots,k_1+k_2\}$ and involves elements
  of the modular S matrices of $SU(2)_k$ which are
      given by
\begin{equation}
  S_{j_1j_2}^{(k)}\ 
=\ \sqrt{\frac{2}{k+2}}\,\sin\frac{\pi}{k+2}(2j_1+1)(2j_2+1)\ \ .
\end{equation}
  We will assume that at least one of the levels is odd in order to
  avoid problems with fixed point resolution.
  (If all levels are even, one needs to consider
  the boundary state
  $|(k_1/4,k_2/4,(k_1+k_2)/4)\rangle$ which is the sum of two
  elementary ones.) It is worth emphasising that among the previous
  boundary states one can recover Cardy states by considering
  boundary labels of the form $(\rho_1,\rho_2,0)$. Since these
  correspond to factorising boundary conditions the transmission has
  to vanish in this case and indeed that is what will be confirmed
  by our general formula below.
\smallskip

  In the overlap \eqref{eq:Overlap} we are interested in, 
  only two sectors of the boundary state \eqref{eq:BS} actually
  contribute, namely the ones with $\mu_1=\mu_2=0$ and $\mu\in\{0,1\}$.
  The Ishibashi states are a sum over a complete set of orthonormal
  states. Hence the relevant contribution
  in the present case is
\begin{equation}
  |(\rho_1,\rho_2,\rho)\rangle
  \ =\ \mc{N}(\rho_1,\rho_2,\rho)
\Biggl[
\Bigl( |v_1\rangle\otimes
   \overline{|v_1\rangle}+|v_2\rangle\otimes\overline{|v_2\rangle}
\Bigr) 
   +\frac{S_{00}^{(k_1+k_2)}S_{\rho1}^{(k_1+k_2)}}
  {S_{\rho0}^{(k_1+k_2)}S_{01}^{(k_1+k_2)}}\,
  |v_3\rangle\otimes\overline{|v_3\rangle}\Biggr]+\cdots
\end{equation}
  with a normalisation $\mc{N}(\rho_1,\rho_2,\rho)$ which will drop
  out in the final expression for the transmission.
  After a straightforward calculation which uses the
  values $c_k=3k/(k+2)$ of the central charges and
  the states defined in eq.\ \eqref{eq:StatesSquare} we
  finally arrive at a nice expression for the
  transmission,
         which turns out to be independent of $\rho_1$ and $\rho_2$,
\begin{equation}
  \Tc_\rho(k_1,k_2)\ 
=\ \frac{2k_1k_2(k_1+2)(k_2+2)}
   {(k_1k_2+k_1+k_2)(k_1+k_2)(k_1+k_2+4)}
  \Biggl\{
 1-\frac{S_{\rho1}^{(k_1+k_2)}S_{00}^{(k_1+k_2)}}
  {S_{\rho0}^{(k_1+k_2)}S_{01}^{(k_1+k_2)}}\Biggr\}\ \ .
\end{equation}
  It is then a natural problem to investigate the level dependence of the
  labels $\rho_{\text{min}}$ and $\rho_{\text{max}}$ which minimise or
  maximise the transmission, respectively. To do so we have to
  extremise the quotient
\begin{equation}
  q(\rho)
\ =\ 
-\frac{S_{\rho1}^{(k_1+k_2)}S_{00}^{(k_1+k_2)}}
      {S_{\rho0}^{(k_1+k_2)}S_{01}^{(k_1+k_2)}}
\ =\ 
-\frac{\sin\frac{3\pi(2\rho+1)}{k_1+k_2+2}}
      {\sin\frac{3\pi}{k_1+k_2+2}}
  \,\frac{\sin\frac{\pi}{k_1+k_2+2}}
         {\sin\frac{\pi(2\rho+1)}{k_1+k_2+2}}
\ =\ 
-\frac{1+2\cos\frac{2\pi(2\rho+1)}{k_1+k_2+2}}
       {1+2\cos\frac{2\pi}{k_1+k_2+2}}\ \ .
\end{equation}
  In the relevant interval for $\rho$ this function has one maximum
  and two minima, namely
\be
\begin{array}{llll}
  \rho_{\text{max}}
  &\ =\ \frac{k_1+k_2}{4}&
  q(\rho_{\text{max}})
  &\ =\ 
  \frac{1}{1+2\cos\frac{2\pi}{k_1+k_2+2}}
  \xrightarrow{\ k_1+k_2\to\infty\ }\frac{1}{3}\\[2mm]
  \rho_{\text{min}}
  &\ \in\  \Bigl\{0,\frac{k_1+k_2}{2}\Bigr\}&
  q(\rho_{\text{min}})
  &\ =\ -1\ \ .
\end{array}
\ee
  It should be noted that the actual maximum of the function $q(\rho)$ 
  is not realised in the range of valid boundary labels if the sum
  of the levels is odd. 
  In that case the maximal transmission is found for
  $\rho = \rho_{\text{max}} \pm1/4$.

\smallskip

  Let us now discuss whether the transmission we calculated satisfies
  all the relevant consistency conditions. First of all, the previous
  calculation provides an explicit proof that the minimal transmission
  for the considered type of defect is indeed zero: $\Tc_{\text{min}}
  =\Tc_{\rho_{\text{min}}}=0$. We notice that vanishing transmission
  is realised precisely for $\rho=0$ (the second value is related to
  0 by the identification rule) where the boundary state \eqref{eq:BS}
  reduces to the product of two Cardy states of $SU(2)_{k_1}$ and
  $SU(2)_{k_2}$, respectively. On the other hand the maximal
  transmission   is given by
\begin{equation}
  \label{eq:TransMax}
  \Tc_{\text{max}}(k_1,k_2)
  \ =\ %
\frac{4k_1k_2(k_1+2)(k_2+2)}{(k_1k_2+k_1+k_2)(k_1+k_2)(k_1+k_2+4)}
  \left\{\begin{array}{cl}
  \frac{1+\cos\frac{2\pi}{k_1+k_2+2}}{1+2\cos\frac{2\pi}{k_1+k_2+2}}
    & \text{ for }k_1,k_2\text{ odd}\\[4mm]
  \frac{1+\cos\frac{\pi}{k_1+k_2+2}}{1+2\cos\frac{\pi}{k_1+k_2+2}}
      & \text{ for }k_1+k_2\text{ odd}\ \ .
       \end{array}\right.
\end{equation}
  It is not entirely obvious from this expression but the transmission 
  turns out to be bounded from above by $1$ as expected, see diagram
  \ref{fig:Trans}\,b).
  Note that one indeed reaches full transmission for $\rho=1/2$
  and $k=1$. In fact this could be expected since these states
  have already been identified in \cite{Quella:2002ct} as permutation
  boundary states of $SU(2)_1\times SU(2)_1$.
\smallskip

\FIGURE[tb]{
  \begin{picture}(190,150)
    \put(-10,0){ \scalebox{.7}{\includegraphics{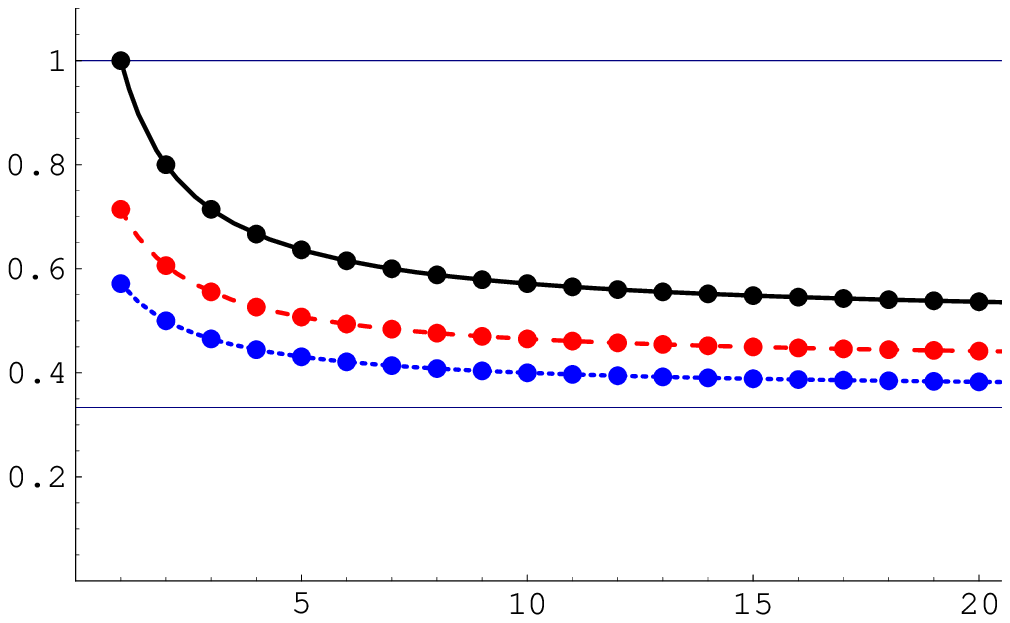}} }
    \put(-10,0){
     \setlength{\unitlength}{.7pt}\put(-59,-16){
     \put(60,205)   {a)}
     \put(90,195)   {\scriptsize $ \Tc(k_1,k_2) $}
     \put(320,15)   {\scriptsize $ k_1 $}
     }\setlength{\unitlength}{1pt}}
  \end{picture}
  \hspace*{1em}
  \begin{picture}(190,150)
    \put(0,0){ \scalebox{.7}{\includegraphics{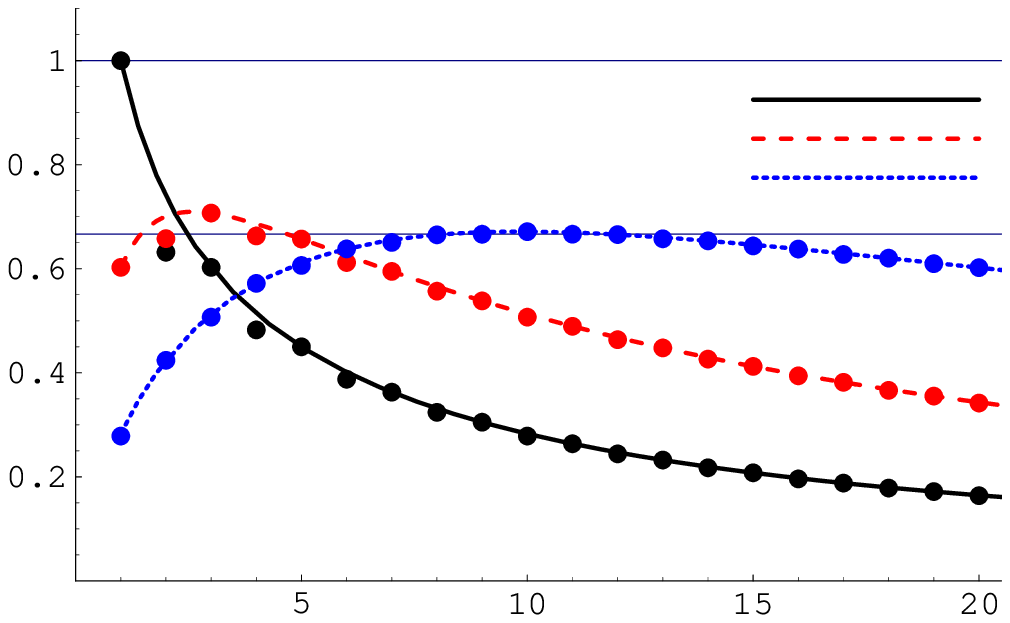}} }
    \put(0,0){
     \setlength{\unitlength}{.7pt}\put(-59,-16){
     \put(60,205)   {b)}
     \put(90,195)   {\scriptsize $ \Tc_{\text{max}}(k_1,k_2) $}
     \put(238,166)   {\scriptsize $ k_2=1 $}
     \put(238,154)   {\scriptsize $ k_2=3 $}
     \put(238,142)   {\scriptsize $ k_2=10 $}
     \put(320,15)   {\scriptsize $ k_1 $}
     }\setlength{\unitlength}{1pt}}
  \end{picture}
  \caption{
  Transmission for specific defects joining $SU(2)_{k_1}$ and
  $SU(2)_{k_2}$ 
  in dependence on $k_1$ for the choices $k_2 = 1,3,10$:
  a) Defect obtained from the common $U(1)$ sub-symmetry (see
  section \ref{sec:SubSym}). b) Defect obtained by embedding
  $SU(2)_{k_1+k_2}$ (see section \ref{sc:Coset}). Here the defect with
  maximal $\Tc$ is considered; the line gives $\Tc_\rho$ for $\rho = 
  \tfrac14(k_1{+}k_2)$. For $k_1+k_2$ even this value is achieved,
  while   for $k_1+k_2$ odd the maximal $\Tc$ occurs for
  $\rho=\tfrac14(k_1{+}k_2{\pm}1)$, so that the corresponding points
  are   slightly below the line.}
  \label{fig:Trans}
}

  The analysis of \eqref{eq:TransMax} simplifies considerably in
  the limit of large levels.
  In that case one can easily derive the limiting value
\begin{equation}
  \Tc_{\text{max}}(k_1,k_2)
  \ \xrightarrow{\ k_1,k_2\gg1\ }\ 
    \frac{8k_1k_2}{3(k_1+k_2)^2}
\end{equation}
  for the maximal transmission for the considered type of defect.
  Note that this value is equal to $2/3$ for $k_1=k_2$ 
  and smaller otherwise.
  In particular it is a monotone decreasing function in each of the
  levels and vanishes when one level is sent to infinity
  while the other one is kept fixed.
\smallskip

  The setting considered here allows us to make contact with the
  generalised permutation branes which have been discussed in
  \cite{Fredenhagen:2005an}.\footnote{We would like to thank Stefan
  Fredenhagen for pointing this out.} Although neither were concrete
  expressions for the boundary states derived nor could the
  precise symmetry preserved be worked out, it was 
  argued that, on geometrical grounds, the diagonal current algebra
  should be preserved. If in the present setup we choose one of the
  levels to be one, e.g.~$k_2=1$, then we preserve the diagonal
  current algebra and, moreover, the additional coset is a Virasoro
  minimal model. Hence in that case we are able to describe {\em all}
  boundary conditions which preserve the diagonal current algebra and,
  in particular, all generalised permutation branes. The maximal 
  transmission in that case is easily worked out
  from eq.~\eqref{eq:TransMax}. 
  It would be interesting to be able to compare this result with
  a calculation genuinely done in the context of generalised
  permutation branes but up to now their algebraic construction seems
  to be beyond reach although recent progress in that direction has
  been reported for products of supersymmetric $SU(2)$ cosets
  \cite{Fredenhagen:2006qw}.
\smallskip

  Before we conclude this section let us briefly investigate
  the action of the topological defects on the boundary conditions
  above. In the $SU(2)_k$ WZW model
  the elementary topological
  defects are labelled by irreducible unitary
  representations
  of the affine Lie algebra. In our situation we have two different
  affine algebras on both sides of the defect. Their action on a
  defect labelled by $(\rho_1,\rho_2,\rho)$ may be expressed in
  terms of the fusion rules of the individual theories as
\begin{equation}
  X_{j_1}\ast D(\rho_1,\rho_2,\rho)\ast X_{j_2}
  \ =\ \sum_{\rho'_1,\rho'_2}\,(N^{(k_1)})_{j_1 \rho_1}^{\rho'_1}\,\,
  (N^{(k_2)})_{j_2 \rho_2}^{\rho'_2}\,D(\rho'_1,\rho'_2,\rho)~~,
\end{equation}
where $2j_i \in \{0,\dots,k_i\}$, $i=1,2$.
  The independence of the transmission amplitude on $\rho_1$
  and $\rho_2$ trivially guarantees that the former is invariant
  under the action of topological defects. 

  One can verify that the conformal defects $D(\rho_1,\rho_2,\rho)$ for
  a fixed value of $\Rc$ are precisely those with a fixed value of
  $\rho$. In particular, all conformal defects (preserving the coset
  symmetry) with a given value of $\Rc$ are generated (in the sense of
  section \ref{sec:RT-top}) by the action of topological defects on
  $D(0,0,\rho)$ for an appropriate $\rho$. 

\sect{Minimal models with rational products}\label{sec:minmod}

In this section we consider the case where $\text{CFT}_1$ and
$\text{CFT}_2$ are (Virasoro) minimal models such that
$\text{CFT}_1 \times \ol{\text{CFT}}_2$ 
is again a minimal model.
Since the central charges add,
this is only possible if at least one of the two is non-unitary.
The fact that the product theory is again a minimal model means
one can find {\em all} conformal boundary conditions, and hence all
conformal defects joining $\text{CFT}_1$ and $\text{CFT}_2$.

There are three products of minimal models which can be analysed
in this way, namely Lee-Yang$\times$Lee-Yang, Lee-Yang$\times$Ising
and Lee-Yang${\times}M_{2,7}$. To see that there cannot be more note
that a product of two minimal models has a chiral symmetry given by
two copies of the Virasoro algebra. Hence to describe it one
has to extend the chiral symmetry of the minimal model used to
describe
the product. The existence of such an extension implies the existence 
of a block-diagonal modular invariant partition function, and these
have been classified \cite{Cappelli:1987xt}. 
The block diagonal cases are the $D_\text{even}$-series for
$M_{p,2(2m+1)}$, the $E_6$ invariant for $M_{p,12}$, and the $E_8$
invariant 
for $M_{p,30}$ (up to the symmetry which interchanges
$p$ and $q$ in $M_{p,q}$). Of these, only 
$D_6$ of $M_{3,10}$, 
$E_6$ of $M_{5,12}$ and 
$E_8$ of $M_{7,30}$ 
involve an extra field of
weight two in the extended symmetry algebra, and these are already
the three cases quoted above.
Furthermore, these three models are also the only cases in which the
sum of the effective central charges of two minimal models is equal to
the effective central charge of a third.

In the following three sections we list the results found for these
three models. The calculations have been done using the category theoretic 
methods of \cite{Fuchs:2002cm} and have been shifted to appendices 
\ref{app:LYxLY}--\ref{app:LYxM27}.

\subsection{Lee-Yang $\times$ Lee-Yang}

The Lee-Yang model is the minimal model $M_{2,5}$ and has central
charge $c={-}\tfrac{22}5$. The product
$M_{2,5} \times M_{2,5}$ is equivalent to the $D_6$-invariant
of $M_{3,10}$. The Kac table of the $c={-}\tfrac{22}5$ Virasoro
algebra contains two irreducible highest
weight representations, which have conformal weight $0$ and
${-}\tfrac15$. We will denote them by $\one$ and $\phi$, respectively.
The fusion product of $\phi$ with itself is $\phi\cdot\phi=\one+\phi$.
It follows from the methods in \cite{Cardy:1989ir,Petkova:2000ip}, which
are valid in general for models with charge-conjugation modular
invariant, that the Lee-Yang model has two conformal boundary
conditions $B_\one$ and $B_\phi$, and two topological defects $X_\one$
and $X_\phi$. The fusion of the topological defects amongst
themselves, and their fusion with the boundary conditions, agrees with
that of the chiral representations, i.e.
\be
  X_\phi \star X_\phi = X_\one + X_\phi
  \quad \text{and}\quad
  X_\phi \star B_\one = B_\phi ~~,~~~
  X_\phi \star B_\phi = B_\one + B_\phi ~.
\ee
The product model on the upper half-plane therefore has at least six
conformal boundary conditions, corresponding to the four factorising
conformal defects $F_{\one\one}$, $F_{\one\phi}$, $F_{\phi\one}$,
$F_{\phi\phi}$ (where the two labels are the boundary condition for
the Lee-Yang model on the upper and lower half-plane, respectively)
and the two topological defects $X_\one$ and $X_\phi$.

In fact, the conformal boundary conditions of the $D_6$-invariant
of $M_{3,10}$ can be labelled by the nodes of the $D_6$ Dynkin diagram
\cite{Behrend:1998fd,Behrend:1999bn}\footnote{
   To be more precise, the boundary conditions are labelled by a pair
   $(x,y)$, where $x$ is a node of the $D_6$ Dynkin diagram and $y$ is an
   odd node of the $A_2$ Dynkin diagram. In the present example there is
   only one choice for $y$ and we have omitted this label, but in the next 
   two examples the $A_n$ label appears explicitly.
   }, 
so the six boundary
conditions mentioned above are already all there is. The relation
between the Lee-Yang and $M_{3,10}$ quantities is computed in 
appendix \ref{app:LYxLY} to be:
\be
  \raisebox{-55pt}{
  \begin{picture}(200,110)
    \put(0,5){ \scalebox{.8}{\includegraphics{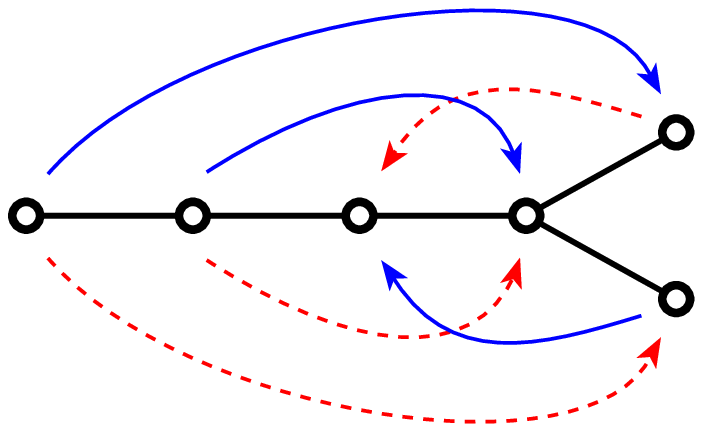}} }
    \put(0,5){
     \setlength{\unitlength}{.8pt}\put(-59,-16){
     \put( 50,64)   {\scriptsize$ F_{\one\one} $}
     \put(104,64)   {\scriptsize$ X_\one $}
     \put(151,64)   {\scriptsize$ F_{\phi\phi} $}
     \put(194,64)   {\scriptsize$ X_\phi $}
     \put(266,97)   {\scriptsize$ F_{\phi\one} $}
     \put(266,49)   {\scriptsize$ F_{\one\phi} $}
     }\setlength{\unitlength}{1pt}}
  \end{picture}}
\labl{eq:M25xM25-D6}
The solid arrows give the result of fusing the corresponding
conformal defect with $X_\phi$ from the left, and the dashed
arrow the result of fusing with $X_\phi$ from the right.
The arrows not included explicitly in the diagram follow
{}from the rule $X_\phi \star X_\phi = X_\one + X_\phi$.

Altogether we see that there are six conformally invariant ways
to join a Lee-Yang model to itself, and they are either purely
reflecting with $\Rc=1$ ($F_{\one\one}$, $F_{\one\phi}$, $F_{\phi\one}$
and $F_{\phi\phi}$) or purely transmitting with $\Tc=1$
($X_\one$ and $X_\phi$).

\subsection{Lee-Yang $\times$ Ising}

The Ising model is the minimal model $M_{3,4}$, and the product
$M_{2,5} \times M_{3,4}$ is equivalent to the $E_6$-invariant of
$M_{5,12}$. The $c=\tfrac12$ Virasoro algebra has three unitary
irreducible highest weight representations. Their conformal weights
are 
$0$, $\tfrac1{16}$ and $\tfrac12$, and we denote them by $\one$,
$\sigma$ and $\eps$, respectively. Their fusion product is given by
$\eps\cdot\eps = \one$, $\sigma \cdot \sigma = 1 + \eps$ and
$\eps\cdot\sigma=\sigma$.

The Ising model has three conformal
boundary conditions, which we denote by $B_\one$, $B_\sigma$ and
$B_\eps$, 
as well as three topological defects, labelled 
$X_\one$, $X_\sigma$ and $X_\eps$. As for the Lee-Yang model, the fusion
of topological defects is given by the fusion of the representations
labelling 
them. For example $X_\sigma \star X_\sigma = X_\one + X_\eps$ or
$X_\sigma \star B_\eps = B_\sigma$.

Consider now the situation where on the upper half-plane we have the
Lee-Yang model and on the lower half-plane the Ising model. Since their
central charges are different, there cannot be any topological defects.
However, there will be six factorising defects, each labelled by a 
pair of boundary conditions, one for the Lee-Yang model and one for the
Ising model: $F_{\one\one}$, $F_{\one\sigma}$, $F_{\one\eps}$,
$F_{\phi\one}$, $F_{\phi\sigma}$, $F_{\phi\eps}$. The total list of
conformal defects is in one-to-one correspondence with
conformal boundary conditions
of the $E_6$-invariant of $M_{5,12}$. There are 12 such boundary
conditions, and they are conveniently described by pairs $(x,y)$,
with $x$ a node of the $E_6$ Dynkin diagram and $y$ an odd node of the
$A_4$ Dynkin diagram. The calculation in appendix \ref{app:LYxIs} yields
the following result for the action of the topological defects on these
12 conformal defects:
\be
  \raisebox{-55pt}{
  \begin{picture}(300,105)
    \put(0,5){ \scalebox{.8}{\includegraphics{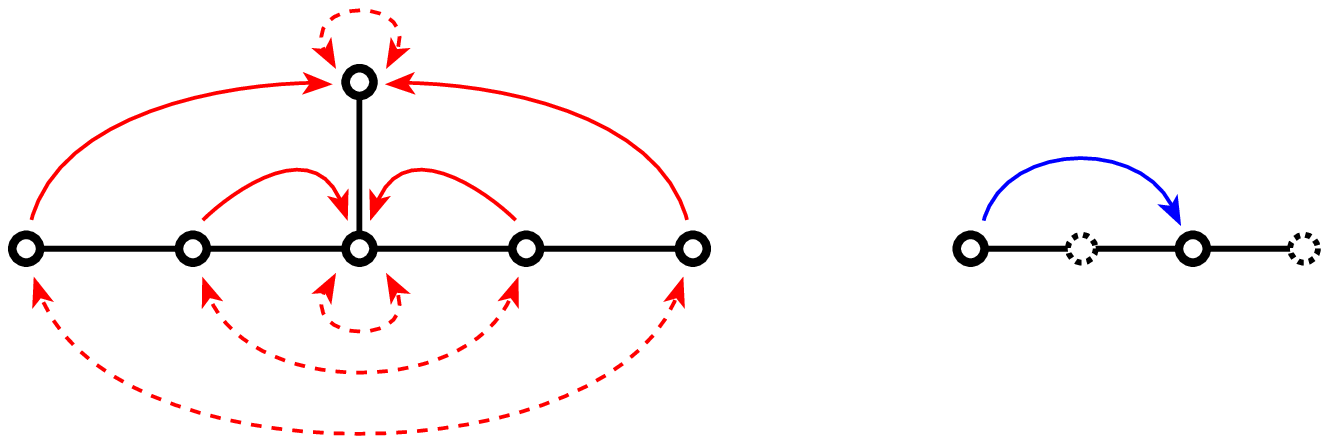}} }
    \put(0,5){
     \setlength{\unitlength}{.8pt}\put(-59,-22){
     \put( 52, 64)  {\scriptsize$ F_{x\one} $}
     \put(264, 64)  {\scriptsize$ F_{x\eps} $}
     \put(169,112)  {\scriptsize$ F_{x\sigma} $}
     \put( 99, 64)  {\scriptsize$ D_{x\one} $}
     \put(215, 64)  {\scriptsize$ D_{x\eps} $}
     \put(139, 83)  {\scriptsize$ D_{x\sigma} $}
     \put(326, 62)  {\scriptsize$ x=\one $}
     \put(395, 62)  {\scriptsize$ x=\phi $}
     }\setlength{\unitlength}{1pt}}
  \end{picture}}
\labl{eq:Ly-Is-nodes}
The dashed arrows in the $E_6$ diagram give the $\Zb_2$-action of the
$X_\eps$ defect, while the solid arrow amounts to the action of
$X_\sigma$. On the $A_2$ diagram, the solid arrow shows the fusion
with $X_\phi$. 
Again, the action of the topological defects on the
12 conformal defects can be reconstructed from those shown explicitly
in \erf{eq:Ly-Is-nodes} together with the fusion product of the
topological  defects. For example, 
$X_\phi \star F_{\one\eps} \star X_\eps = F_{\phi\one}$, or
$X_\phi \star D_{\phi\sigma} \star X_\sigma = 
D_{\one\one} + D_{\phi\one} + D_{\one\eps} + D_{\phi\eps}$.

For the six factorising defects, $\Rc(F_{xy})=1$, while for
the remaining six one finds (see appendix \ref{app:LYxIs})
\be
  \Rc(D_{xy}) = \frac{2841+440\sqrt{3}}{1521} = 2.3689.. 
  \quad,\quad
  \Tc(D_{xy}) = 1-\Rc = -1.3689.. \quad.
\labl{eq:M25-M34-refl-trans}
This provides an example that $\Rc$ and $\Tc$ can lie outside the
interval $[0,1]$ when non-unitary models are involved.

\subsection{Lee-Yang $\times \,{\bf M_{2,7}}$}

The product theory $M_{2,5} \times M_{2,7}$ is given by the
$E_8$-invariant of $M_{7,30}$. The minimal model $M_{2,7}$ has
central charge $c={-}\tfrac{68}7$, and there are three
irreducible Virasoro highest weight representations in the Kac table.
These have conformal weights $0$, ${-}\tfrac27$, ${-}\tfrac37$
and will be labelled $\one$, $\alpha$, $\beta$, respectively. 
The fusion rules are 
$\alpha \cdot \alpha = \one + \beta $, 
$\alpha \cdot \beta = \alpha + \beta$ and
$\beta \cdot \beta = \one + \alpha + \beta$.
There are three conformal boundary conditions $B_y$
and three topological defects $X_y$ for $M_{2,7}$, with
$y$ taking values in $\{\one,\alpha,\beta\}$. The action of
the topological defects is again given by the fusion rules
of $\alpha$ and $\beta$.

The analysis of the conformal defects which can join $M_{2,5}$ and
$M_{2,7}$ is similar to the previous two minimal model examples.
There will be six factorising defects and no topological ones. 
The product model, described as the $E_8$-invariant of $M_{7,30}$, 
has 24 conformal boundary conditions. They are labelled by pairs
$(x,y)$, with $x$ a node of the $E_8$ Dynkin diagram and
$y$ an odd node of the $A_6$ Dynkin diagram. 

The action of the topological
defects of $M_{2,5}$ and $M_{2,7}$ on the conformal
defects corresponding to these boundary conditions is given
by:
\be
  \raisebox{-10pt}{
  \begin{picture}(400,100)
    \put(-10,5){ \scalebox{.8}{\includegraphics{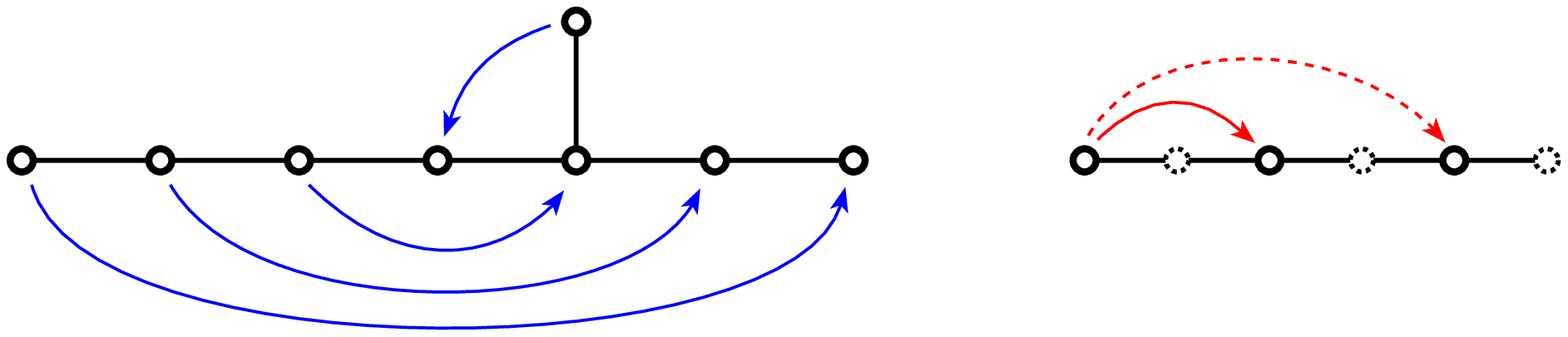}} }
    \put(-10,5){
     \setlength{\unitlength}{.8pt}\put(-59,-22){
     \put( 62, 93)  {\scriptsize$ F_{\one y} $}
     \put(110, 93)  {\scriptsize$ D_{1y} $}
     \put(158, 93)  {\scriptsize$ D_{2y} $}
     \put(206, 64)  {\scriptsize$ D_{3y} $}
     \put(267, 93)  {\scriptsize$ D_{4y} $}
     \put(271,130)  {\scriptsize$ D_{6y} $}
     \put(302, 93)  {\scriptsize$ D_{5y} $}
     \put(350, 93)  {\scriptsize$ F_{\phi y} $}
     \put(427, 64)  {\scriptsize$ y=\one $}
     \put(492, 64)  {\scriptsize$ y=\beta $}
     \put(558, 64)  {\scriptsize$ y=\alpha $}
     }\setlength{\unitlength}{1pt}}
  \end{picture}}
\labl{eq:M25xM27-E8}
The arrows in the $E_8$ picture give the result of fusing with
$X_\phi$.\footnote{
  We note that, intriguingly,
  the relation between nodes resulting from
  the action of $X_\phi$ as given for the $D_6$-diagram in
  \erf{eq:M25xM25-D6} and for the $E_8$-diagram in \erf{eq:M25xM27-E8}
  has also been observed in \cite{Fring:2005vg} 
  in an apparently unrelated context.
}
The dashed arrow in the $A_6$ diagram gives the fusion
with $X_\alpha$ and the solid arrow the fusion with $X_\beta$.

The exact reflection and transmission coefficients are somewhat
cumbersome to spell out because of the fractions arising
{}from the central charges, and it is much briefer to give the
coefficients $\X$ instead, which are related to $\Rc$ and $\Tc$
via \erf{eq:RT-via-omega}. The computations can be found
in appendix \ref{app:LYxM27}, and the results are,
for $y \in \{\one,\alpha,\beta\}$,
\be\begin{array}{ll}\displaystyle
  \X(F_{\one y}) = \X(F_{\phi y}) = 1 \etb \Rightarrow ~~\Rc = 1
  \enl
  \X(D_{1y}) = \X(D_{5y}) 
  = \cos\!\big(\pi\tfrac{17}{30}\big) 
 / \cos\!\big(\pi\tfrac{7}{30}\big) 
  \etb \Rightarrow ~~\Rc = 0.4508..
  \enl
  \X(D_{2y}) = \X(D_{4y}) 
  = {-}\sin\!\big(\pi\tfrac{7}{30}\big) 
 / \sin\!\big(\pi\tfrac{17}{30}\big) 
  \etb \Rightarrow ~~\Rc =  0.2773..
  \enl
  \X(D_{3y}) = \X(D_{6y}) 
  = \cos\!\big(\pi\tfrac{1}{30}\big) 
 / \cos\!\big(\pi\tfrac{11}{30}\big) 
  \etb \Rightarrow ~~\Rc =  1.6201..\ \ .
\eear\labl{eq:M27-Xb-values}
We see that also in this example there are six conformal defects
where $\Rc$ (and hence also $\Tc$) lies outside the interval $[0,1]$.

\sect{Conclusions and outlook}

In this article we proposed two new quantities $\Rc$ and
$\Tc$ which may be used to characterise conformal defects,
and we determined their value for a variety of critical defect
systems. In the case of the free boson 
                     and the free fermion
they reduce to the
reflection and transmission probability, respectively.
There are, however, indications that the interpretation
as reflection and transmission is valid rather generically.
First of all, our definition leads to the values $\Tc=0$
and $\Tc=1$ for factorising and topological defects,
respectively, whose physical interpretation as fully reflective
and fully transmissive is obvious. Moreover, they satisfy
$\Tc+\Rc=1$ and for unitary theories
take values in the interval $[0,1]$, at
least for all the examples we considered.
We also showed that, in contrast to the $g$-factors
\cite{Affleck:1991tk}, the
quantities $\Rc$ and $\Tc$ are constant under the action
of topological defects.
On the other hand, as seen for the free boson and the Ising model,
while $g$ is constant under (truly) marginal 
deformations of the defect, $\Rc$ and $\Tc$ are typically not.

Given these appealing features it is then natural
to ask whether the specified bounds on $\Rc$ and
$\Tc$ in unitary models can be shown to hold
rigorously or whether they can finally even be improved.
To this end note that if, in the notation of \erf{eq:rijdef}, the
two ratios $\cev 0 \bar L^i_2 L^i_2 \vec b/\cev 0 \,b\, \rangle$ are
non-negative, i.e.\ if $R_{11}\ge0$ and $R_{22}\ge 0$, then by
\erf{eq:R-via-omb} we have ${-}\X_b \le 
\min\big(\tfrac{c_1}{c_2},\tfrac{c_2}{c_1}\big)$. 
This results in the bounds
\be
  \Tc\ \leq\ \frac{2\min(c_1,c_2)}{c_1+c_2} \qquad \text{and} \qquad
  \Rc\ \geq\ \frac{|c_1-c_2|}{c_1+c_2} ~~.
\labl{eq:bounds}
These bounds hold in all unitary models we considered. In the class of
examples treated in section \ref{sec:SubSym} they can even be
saturated. This happens when the chiral algebra of CFT$_1$ (say) is
contained  in that of CFT$_2$, and the defect is taken to be
transmissive for all fields in this chiral algebra, so that $c=c_1$ in
\erf{eq:SubTrans}. However, we have no general proof of the assertion
\erf{eq:bounds}. Another observation worth mentioning is that the
bound $\Rc \ge 0$ holds even for the non-unitary minimal models
considered in section \ref{sec:minmod}. 

Independent of such structural considerations the
quantities $\Rc$ and $\Tc$ may be
useful in the analysis of RG flows in defect systems.
There are three ways to deform two conformal field theories separated  
by a defect: one can turn on a perturbation in the
bulk of either CFT, and one can perturb the defect itself. 
In the last case the behaviour of the bulk theories
will be unaffected. Yet, in all cases the defect will start
to flow and it is an interesting question to ask what the new
infrared fixed point is. The
analysis of this problem is usually very hard and requires
the use of numerical methods to arrive at concrete predictions
\cite{Dorey:1997yg,Dorey:1999cj}.
For pure defect flows there do exist fairly
general conjectures \cite{Fredenhagen:2002qn,Fredenhagen:2003xf}
for the end point of the flows. We expect the quantities $\Rc$
and $\Tc$ to be a useful tool to determine the
actual fixed point in numerical calculations.

One of the initial motivations for our study of transmission
amplitudes was to improve our understanding of so-called
generalised permutation branes.
These are conjectural non-factorising boundary conditions
in product CFTs whose constituents are structurally very close,
such as minimal models with distinct central charge, or WZW
(or coset) models based on the same group but at different
levels. The common feature of these models is that
they come in a multi-parameter family and that on a diagonal
subset where two or more parameters agree one has a permutation
symmetry. The latter may then be used to construct topological defects.
It appears reasonable to assume that, when slightly moving away
{}from the diagonal, these defects will not suddenly completely change
their nature, in particular in the semi-classical regime where
the parameters (e.g.\ the levels) are large and geometric reasoning
applies. This picture has been confirmed in \cite{Fredenhagen:2005an} 
in the case of product groups. Generalised permutation branes
are also known to exist in $N=2$ topological conformal field
theories \cite{Brunner:2005fv,Caviezel:2005th}. The only
rigorous CFT construction has been performed in
\cite{Fredenhagen:2006qw} and it is restricted to very few
and rather special examples.

Despite all these activities we still miss a satisfactory answer
at the moment to the question of what qualifies a defect
for being associated to a `generalised permutation brane'.
Lacking any additional algebraic input, e.g.\ from symmetry,
one could be tempted to call generalised permutation branes
those branes in the families just described whose transmission
can be brought arbitrarily close to one, for instance for
large values of the parameters close to the diagonal.
                   One possibility is
that away from the
diagonal among all possible defects those related
to generalised permutation branes realise the maximal
transmission coefficient that can occur for this
specific choice of parameters.

In fact our investigation of non-unitary minimal models
in this paper was precisely motivated by this reasoning.
Indeed, in these models {\em all} the defects are known,
and hence also those which are possibly connected to
generalised permutation branes. In this context it
is worth recalling that the Ising model possesses a relevant
perturbation by an imaginary magnetic field which lets it
flow to the Lee-Yang edge singularity. If we now consider
the Ising model with a topological defect and switch on
an imaginary magnetic field in the upper half-plane we
end up with a possibly non-factorising defect which
separates a Lee-Yang phase {}from an Ising phase and which
might possess an interpretation as a generalised permutation brane.
Perturbing first the upper half-plane and then the lower half
plane gives a sequence of flows from
Ising$\times$Ising over Lee-Yang$\times$Ising to 
Lee-Yang$\times$Lee-Yang, and these are precisely the models
where we determined the transmission for a complete set
of conformal defects. Since the transmission and reflection
do not seem to satisfy reasonable bounds in these non-unitary
models and since the combined bulk/defect flows are
beyond our control so far, we could not draw any conclusions regarding
the interpretation of certain boundary states as generalised
permutation branes. Nevertheless it appears to be a
promising project to reanalyse the situation in a similar
setup with unitary models, possibly in the semi-classical
regime where perturbation theory is at hand.

\acknowledgments

It is a great pleasure to thank Zoltan Bajnok, 
Pasquale Calabrese, Jean-Sebastien Caux, Patrick Dorey,
Stefan Fredenhagen and Matthias Gaberdiel  
for useful discussions and J\"urgen Fuchs, Andreas
Recknagel and Christoph Schwei\-gert 
for valuable comments on the manuscript.
Also, we are most grateful to the referee for pointing out 
references \cite{Friedan:2005bz,Friedan:2005ca} to us.
TQ acknowledges the hospitality of the KITP in Santa Barbara during the
completion of this work.
This work was supported in part by the EU Research Training Network
grants  `Euclid', contract number HPRN-CT-2002-00325, `Superstring
Theory',  contract number MRTN-CT-2004-512194, by the PPARC rolling
grant PP/C507145/1, and by the National Science Foundation
under Grant no.\ PHY99-07949.
Until September 2006 TQ was funded by a
PPARC postdoctoral fellowship under reference PPA/P/S/2002/00370.
IR receives partial support from the EPSRC First Grant
EP/E005047/1.

\appendix

\sect{Reflection and Transmission via $\X_b$}\label{app:RT-omega}

In the following we give some details regarding the derivation
of \erf{eq:R-via-omb}. Define the (anti-) holomorphic fields $W$ and
$\ol W$ as
\be
  W = c_2 T_1 - c_1 T_2 ~~,~~
  \ol W = c_2 \ol T_1 - c_1 \ol T_2 ~~,
\labl{eq:W-Wbar-def}
for $c_1$, $c_2$ the central charges of $\text{CFT}_1$ and
$\text{CFT}_2$. One can check that $W$ and $\ol W$ are primary
with respect to $T^\text{tot}$ and $\ol T{}^\text{tot}$,
and that the OPE of $W$ with itself reads
\be
  W(z)W(w) = 
  \frac{\tfrac12 c_1 c_2 (c_1{+}c_2)}{(z-w)^4}
  + \Big( \frac{2}{(z{-}w)^2} + \frac{1}{z{-}w} \frac{\partial}{\partial w}
  \Big) \big( (c_2{-}c_1)W(w) + c_1 c_2 T^\text{tot}(w)\big)
  + \cdots
\ee
For a boundary state $|b\rangle$ we have
\be
  \langle 0 | W_2 \ol L{}^\text{tot}_2 |b\rangle = 0
  \qquad \text{and} \qquad
  \langle 0 | L{}^\text{tot}_2 \ol W_2 |b\rangle = 0 ~~,
\labl{eq:WTtot=0}
since e.g.\ $\langle 0 | W_2 \ol L{}^\text{tot}_2 |b\rangle 
= \langle 0 |  W_2 L{}_{-2}^\text{tot} |b \rangle$ and
$\langle 0 |  W_2 L{}_{-m}^\text{tot} = 0$ for $m>0$ as $W$ is
primary. 
To proceed we need to assume that
\be
  c_1 \neq 0 ~~,~~~
  c_2 \neq 0 \quad \text{and} \quad
  c_1{+}c_2 \neq 0 ~~.
\labl{eq:centc-assume}
Using the conditions \erf{eq:WTtot=0}, written out 
in terms
of the $L^{1,2}_m$ and $\ol L{}^{1,2}_m$, it is
straightforward to verify the identities
\bea
   \langle 0| L^1_2 \ol L{}^1_2 |b\rangle = \frac{1}{(c_1{+}c_2)^2} 
   \Big( c_1^{\,2}\, 
   \langle 0| L^\text{tot}_2 \ol L{}^\text{tot}_2 |b\rangle +
   \langle 0| W_2 \ol W{}_2 |b\rangle \Big)
   \enl
   \langle 0| L^2_2 \ol L{}^2_2 |b\rangle = \frac{1}{(c_1{+}c_2)^2} 
   \Big( c_2^{\,2}\, 
   \langle 0| L^\text{tot}_2 \ol L{}^\text{tot}_2 |b\rangle +
   \langle 0| W_2 \ol W{}_2 |b\rangle \Big)
   \enl
   \langle 0| L^1_2 \ol L{}^2_2 |b\rangle = 
   \langle 0| L^2_2 \ol L{}^1_2 |b\rangle = 
   \frac{1}{(c_1{+}c_2)^2} 
   \Big( c_1 c_2 \, 
   \langle 0| L^\text{tot}_2 \ol L{}^\text{tot}_2 |b\rangle -
   \langle 0| W_2 \ol W{}_2 |b\rangle \Big)
\eear\labl{eq:TT_via_WW}
Of course we can also write
$\langle 0| L^\text{tot}_2 \ol L{}^\text{tot}_2 |b\rangle
= \langle 0| L^\text{tot}_2 L^\text{tot}_{-2} |b\rangle
= \tfrac12 (c_1{+}c_2) \langle 0 |b\rangle$.
To obtain \erf{eq:R-via-omb} we just need to introduce
\be
  \X_b = \frac{2}{c_1 c_2 (c_1{+}c_2)} \cdot
  \frac{\langle 0| W_2 \ol W{}_2 |b\rangle}{\langle 0| b\rangle} ~~,
\labl{eq:omegab-def}
use this to replace $\langle 0| W_2 \ol W{}_2 |b\rangle$ in
\erf{eq:TT_via_WW} and divide both sides by $\langle 0 |b\rangle$.

Note that for a factorising boundary condition
$|b\rangle  = |b_1\rangle \oti |b_2\rangle$, the inner product
$\langle 0| L^1_2 \ol L{}^2_2 |b\rangle$ will be zero, i.e.\ the entry
$R_{12}$ of the matrix $R$ in \erf{eq:R-via-omb} is zero.
Together with our assumptions 
\erf{eq:centc-assume} on the central charges this implies
\be
  \X_b = 1
  \quad \text{for~factorising~defects}~.
\labl{eq:Xb=1-for-fact}
On the other hand, for a permutation boundary condition (corresponding 
to a topological defect in the unfolded model), we have
$\langle 0| L^1_{2} \ol L{}^1_{2} |b\rangle
= \langle 0| L^1_{2} L^2_{-2} |b\rangle
= 0$. Since furthermore the existence of topological defects in the
unfolded 
model implies $c_1=c_2$, together with \erf{eq:R-via-omb} we obtain
\be
  \X_b = -1
  \quad \text{for~topological~defects}~.
\labl{eq:Xb=-1-for-top}

Finally, instead of the exterior of the unit circle, 
consider the folded CFT on the
upper half-plane with boundary condition $b$. Since the boundary
condition is conformal, we have $T^\text{tot}(x) = \ol
T{}^\text{tot}(x)$ 
for points $x$ on the real axis. {}From the definition 
\erf{eq:W-Wbar-def} it is then not difficult
to see that
\be
  W(x) = \ol W(x) \quad
  \text{for~all~real~$x$~if~and~only~if~$b$~is~factorising} ~.
\labl{eq:W=Wbar-equiv-fact}

\sect{Summary on states in the coset decomposition}\label{app:coset}

  In this appendix we present the bulky formulas for the states that
  are 
  relevant in section \ref{sc:Coset} of this note. We will use the
  spin basis in which the generators of the affine $SU(2)_k$ satisfy
\begin{equation}
  [E_m^+,E_n^-]\;=\;k\,\delta_{m+n,0}+2H_{m+n}\ ,\ %
  [H_m,E_n^\pm]\;=\;\pm E_{m+n}^\pm\ ,\ %
  [H_m,H_n]\;=\;\frac{k}{2}\,\delta_{m+n,0}\ \ .
\end{equation}
  The affine highest weight states $|j\rangle$ are introduced as
  usual by demanding
\begin{align}
  \label{eq:CurAlg}
  H_0|j\rangle&\ =\ j|j\rangle\;,&
  E_0^+|j\rangle&\ =\ 0\;,&
  E_n^\pm|j\rangle&\ =\ H_n|j\rangle\ =\ 0\ \ \text{ for }n>0\ \ .
\end{align}
  With these definitions it is then a straightforward exercise to
  come up with the following set of orthonormalised singlet states
\begin{equation}
  \label{eq:CandVectors}
  \begin{split}
    |v_1\rangle
&\ =\ \sqrt{\frac{2}{c_{k_1+k_2}}}\,|(0,0,0)\rangle
      \otimes L_{-2}|0\rangle\\[2mm]
    |v_2\rangle
&\ =\ \sqrt{\frac{2}{c_{k_1}+c_{k_2}-c_{k_1+k_2}}}
    \,L_{-2}|(0,0,0)\rangle\otimes|0\rangle\\[2mm]
    |v_3\rangle
&\ =\ \frac{1}{4\sqrt{3(k_1+k_2+4)}}\,|(0,0,1)\rangle
    \otimes\bigl(4E_{-1}^--H_{-1}E_0^--2E_{-1}^+E_0^-E_0^-\bigr)|1\rangle
  \end{split}
\end{equation}
  in the decomposed theory \eqref{eq:Deco}
  which are relevant for the description of the
  states $L_{-2}^{i}\,|0\rangle$ appearing in
  eq.~\eqref{eq:rijdef}.
\smallskip

  Using the concrete form of the Sugawara and GKO energy momentum
  tensors and the definition of the state
  $|(0,0,1)\rangle\otimes|1\rangle$   (see \eqref{eq:SpecState}) 
  one can easily show that the linear combinations
  relevant to a solution of our problem are given by
\begin{eqnarray}
    L_{-2}^{1}\,|0\rangle
    &=& \sqrt{\frac{3k_1^2}{2(k_1+k_2)(k_1+k_2+2)}}\,\vec{v_1}
          +\sqrt{\frac{3k_1k_2(k_2+2)}{2(k_1+2)(k_1+k_2+2)(k_1+k_2+4)}}
          \,|v_2\rangle
\nonumber\\[2mm]
    &&+\sqrt{\frac{3k_1k_2}{(k_1+k_2)(k_1+k_2+4)}}\,\vec{v_3}
  \nonumber\end{eqnarray}\begin{eqnarray}
    L_{-2}^{2}\,|0\rangle
    &=& \sqrt{\frac{3k_2^2}{2(k_1+k_2)(k_1+k_2+2)}}\,\vec{v_1}
          +\sqrt{\frac{3k_1k_2(k_1+2)}{2(k_2+2)(k_1+k_2+2)(k_1+k_2+4)}}
      \,\vec{v_2}
\nonumber\\[2mm]
    &&-\sqrt{\frac{3k_1k_2}{(k_1+k_2)(k_1+k_2+4)}}\,|v_3\rangle
\nonumber\\[2mm]
    L_{-2}^{\text{tot}}\,|0\rangle
    &=& |(0,0,0)\rangle\otimes
    L_{-2}|0\rangle+L_{-2}|(0,0,0)\rangle\otimes|0\rangle
\nonumber\\[2mm]
    &=& \sqrt{\frac{c_{k_1+k_2}}{2}}\,|v_1\rangle
+ \sqrt{\frac{c_{k_1}+c_{k_2}-c_{k_1+k_2}}{2}}\,|v_2\rangle\ \ .
  \label{eq:States}
\end{eqnarray}
  Identical relations hold for the anti-holomorphic sector. Combining
  both sectors we thus find
\begin{equation}
  \label{eq:StatesSquare}
  \begin{split}
    L_{-2}^{1}|0\rangle\otimes\bar{L}_{-2}^{1}\overline{|0\rangle}
    &\ =\ \frac{3k_1^2}{2(k_1+k_2)(k_1+k_2+2)}\,|v_1\rangle
     \otimes\overline{|v_1\rangle}\\[2mm]
    &\qquad+\frac{3k_1k_2(k_2+2)}{2(k_1+2)(k_1+k_2+2)(k_1+k_2+4)}
    \,|v_2\rangle\otimes\overline{|v_2\rangle}\\[2mm]
    &\qquad+\frac{3k_1k_2}{(k_1+k_2)(k_1+k_2+4)}\,|v_3\rangle
    \otimes\overline{|v_3\rangle}
           +\text{mixed contributions}\\[2mm]
    L_{-2}^{2}|0\rangle\otimes\bar{L}_{-2}^{2}\overline{|0\rangle}
    &\ =\ \frac{3k_2^2}{2(k_1+k_2)(k_1+k_2+2)}\,|v_1\rangle
\otimes\overline{|v_1\rangle}\\[2mm]
    &\qquad+\frac{3k_1k_2(k_1+2)}{2(k_2+2)(k_1+k_2+2)(k_1+k_2+4)}
\,|v_2\rangle\otimes\overline{|v_2\rangle}\\[2mm]
    &\qquad+\frac{3k_1k_2}{(k_1+k_2)(k_1+k_2+4)}\,|v_3\rangle
\otimes\overline{|v_3\rangle}
           +\text{m.c.}\\[2mm]
    L_{-2}^{1}|0\rangle\otimes\bar{L}_{-2}^{2}\overline{|0\rangle}
    &\ =\ \frac{3k_1k_2}{2(k_1+k_2)(k_1+k_2+2)}\,|v_1\rangle
\otimes\overline{|v_1\rangle}\\[2mm]
    &\qquad+\frac{3k_1k_2}{2(k_1+k_2+2)(k_1+k_2+4)}\,|v_2\rangle
\otimes\overline{|v_2\rangle}\\[2mm]
    &\qquad-\frac{3k_1k_2}{(k_1+k_2)(k_1+k_2+4)}\,|v_3\rangle
\otimes\overline{|v_3\rangle}
           +\text{m.c.}\\[2mm]
    L_{-2}^{\text{tot}}\,|0\rangle\otimes\bar{L}_{-2}^{\text{tot}}
\,\overline{|0\rangle}
    &\ =\ \frac{c_{k_1+k_2}}{2}\,|v_1\rangle\otimes\overline{|v_1\rangle}
          +\frac{c_{k_1}+c_{k_2}-c_{k_1+k_2}}{2}\,|v_2\rangle
\otimes\overline{|v_2\rangle}
          +\text{m.c.}\ \ .
  \end{split}
\end{equation}
  The mixed contributions ``m.c.'' have vanishing overlap
  with Ishibashi states  and have therefore not been spelled out.

\sect{Category theoretic calculations}

In this appendix we will briefly describe the calculations
behind the results stated in section~\ref{sec:minmod}. We will make
use of the `topological field theory approach to rational conformal
field theory', or TFT-approach for short, developed in 
\cite{Felder:1999mq,Fuchs:2002cm,Fuchs:2004xi,Fjelstad:2005ua}. 
Rather than reviewing this formalism,
we highlight a few points in the next section, and also provide
references to the relevant parts of these papers where more details
can be found.

\subsection{Preliminaries}

In the TFT approach one starts from a rational chiral algebra
$\Vc$ (a conformal vertex algebra) and considers its representation
category $\Cc = \text{Rep}(\Vc)$. For Virasoro minimal models, 
$\Vc$ is the Virasoro vertex algebra at the corresponding central charge,
and the simple objects in $\Cc$ are labelled by entries of the Kac table
(modulo its $\Zb_2$-symmetry).

One of the properties one demands of a rational CFT is that the
category $\Cc$ should be modular, i.e.\ (roughly) a $\Cb$-linear
semi-simple braided tensor category, 
where the braiding obeys a certain non-degeneracy
condition (see e.g.\ \cite[app.\,A.1]{Fjelstad:2005ua} for details and
references).  
Given a modular tensor
category $\Cc$, one can construct a three-dimensional topological field
theory (TFT), which has the chiral CFT described by $\Vc$ as its boundary
degrees of freedom (see \cite[app.\,A.2]{Fjelstad:2005ua} and
references therein). 

A full conformal field theory that includes $\Vc$ in its chiral symmetries
can now be described by a symmetric special Frobenius algebra $A$ in the
category $\Cc$ (see \cite[app.\,B]{Fjelstad:2005ua} for a summary). 
Let us denote this conformal field theory 
by $\text{CFT}_A$.
Properties of $\text{CFT}_A$, like boundary conditions
(preserving $\Vc$ with trivial gluing automorphism) or topological
defects transparent to the fields in $\Vc$, are related to natural 
quantities obtained from $A$, like modules and bimodules, respectively
(see \cite[sect.\,4.4]{Fuchs:2002cm}). 
Structure constants of the CFT are expressed as invariants of
framed links in a three-manifold; the value of these invariants
is computed via the TFT defined by $\Cc$ \cite{Fuchs:2004xi}.

\subsubsection{Induced modules}

As already mentioned, modules of $A$ describe boundary conditions of
$\text{CFT}_A$, and simple modules correspond to elementary boundary
conditions. 
The simplest class of $A$-modules are the induced modules. Given an
object $U$ of $\Cc$, the induced module $\inda(U)$ is simply $A \oti
U$ with the action of $A$ given by the multiplication of $A$. One can
show that in the present setting, every simple module is a submodule
of an induced 
module (see \cite[sect.\,4.3]{Fuchs:2002cm} for details and references).
So to find all simple $A$-modules we have to decompose all
induced modules based on simple objects of $\Cc$.
For two $A$-modules $M,N$
denote by $\Hom_A(M,N)$ the subspace of morphisms in $\Cc$
{}from $M$ to $N$, which commute with the action of $A$. The following
property 
is very useful when decomposing induced modules,
\be
  \Hom_A( \inda(U), M ) \cong \Hom(U, M) ~~,
\labl{eq:induced-homs}
where the second $\Hom$-space denotes morphisms in $\Cc$.
In particular, if $\Hom_A( \inda(U)$, $\inda(U) ) \cong 
\Hom(U,A \oti U)$ is one-dimensional, then
$\inda(U)$ is simple, and if $\Hom(A \oti U, U)$ is two-dimensional,
then 
$\inda(U)$ decomposes into two non-isomorphic simple $A$-modules.

We will denote the boundary condition of $\text{CFT}_A$ labelled by
an $A$-module $M$ by $B_{M}$.

\subsubsection{Local modules and factorising defects}

In \erf{eq:W=Wbar-equiv-fact} we have seen that a boundary condition 
of the folded model
(or a defect line in the unfolded model) is factorising if and only if
$W = \ol W$ on the real axis. In the minimal model examples treated below,
the algebra $A$ describes the extension of the symmetry from the
Virasoro algebra at central charge $c_1{+}c_2$ to the product of the
two Virasoro (vertex) algebras with central charges $c_1$ and
$c_2$. In particular, this extension 
contains the primary field $W$.
In terms of the representation category $\Cc$ this means that
the representation $U_W$ of the chiral algebra which contains the
field $W$, is a subobject of $A$.
Also, if $A$ describes an extended symmetry, it will be a 
{\em commutative} special symmetric Frobenius algebra. So from here on
we will assume that $A$ is commutative. Note that from the definition
given above, $A$, seen as a left module over itself, is local.

Recall that $\Cc$ is in particular a braided tensor category; the
braiding is denoted by $c_{UV} \in \Hom(U \oti V,V \oti U)$.
Let $M$ be an $A$-module and let $\rho_M \in \Hom(A\oti M,M)$ be the
representation morphism (see \cite[app.\,A.4]{Fjelstad:2005ua}
for details). The module $M$ 
is called {\em local} iff $\rho_M \circ c_{MA} \circ c_{AM} = \rho_M$.
One can show that a simple module $M$ is local if and only if
all irreducible representations of the chiral algebra entering
$M$ have the same conformal weight modulo $\Zb$ (i.e.\ if the
twist on $M$ is a multiple of the identity, see 
e.g.\ \cite[corollary 3.18]{Frohlich:2003hm} for a proof).

As explained in \cite[sect\,3.3]{Fuchs:2004xi}, 
the multiplicity space of bulk fields of $\text{CFT}_A$ 
transforming in the representation $U_i \times U_j$ of the left/right
copy of the chiral symmetry is isomorphic to a certain morphism
space $\Hom_{A|A}(U_i \otimes^+ A \otimes^- U_j,A)$. If we pick an
embedding $e_W \in \Hom(U_W,A)$, we can take the chiral field $W$
to correspond to the morphism 
$\phi_W = m \circ (e_W \oti \id_A) \in 
\Hom_{A|A}(U_W \otimes^+ A \otimes^- \one,A)$ (this uses that $A$ is
commutative), and similarly the anti-chiral field $\ol W(z)$
corresponds to 
$\phi_{\ol W} = m \circ (\id_A \oti e_W) \in 
\Hom_{A|A}(\one \otimes^+ A \otimes^- U_W,A)$.

{}From the ribbon graph representation \cite[eqn.\,(4.15)]{Fuchs:2004xi}
of an upper half-plane correlator
involving $W$ or $\ol W$, and a boundary condition $B_M$ labelled by the
module $M$, it is not too difficult to deduce that
\be
  M ~~\text{local}
  \qquad \Rightarrow \qquad 
  B_M~~\text{is~factorising~boundary~condition}~~.
\ee
The converse holds if $A$ is generated by the subobject $U_W$ in
an appropriate sense, but we will not need this here. Note also that
since $A$ is a local module over itself (we assumed $A$ to be commutative),
it describes a factorising boundary condition.

\subsubsection{Computing the coefficient $\X$}

Denote by $|M\rangle$ the boundary state described by
an $A$-module $M$, and by $|U\rrangle$ the Ishibashi
state belonging to the simple object $U$ (which is an
irreducible representation of the chiral algebra). As usual
one can write the boundary state as a sum of Ishibashi states,
and accordingly there are constants $C(U_W;M)$ such that
\be
  \langle 0| W_2 \ol W{}_2 |M \rangle
  = C(U_W;M) \langle 0| A \rangle \cdot
  \langle 0| W_2 \ol W{}_2 |U_W \rrangle ~~,
\labl{eq:tft-bndstate}
where we chose to normalise the coefficients
$C(U_W;M)$ relative to $\langle 0| A \rangle$, the inner product 
of the out-vacuum with the boundary state corresponding to $A$.
In the TFT approach, $C(U_W;M)$ can be expressed
as the invariant of a ribbon graph. If, similar to the previous
section, we take the morphism
in $\Hom_{A|A}(U_W \otimes^+ A \otimes^- U_W,A)$ corresponding
to the non-chiral field $W \ol W$ to be $\phi_{W\ol W} = 
m \circ (m \oti \id_A) \circ (e_W \oti \id_A \oti e_W)$,
then according to \cite[eqn.\,(4.20)]{Fuchs:2004xi} the relevant 
invariant is
\be
  C(U_W;M) = \frac{1}{\dim(A)} ~~~~
  \raisebox{-60pt}{
  \begin{picture}(90,120)
    \put(0,0){ \scalebox{.7}{\includegraphics{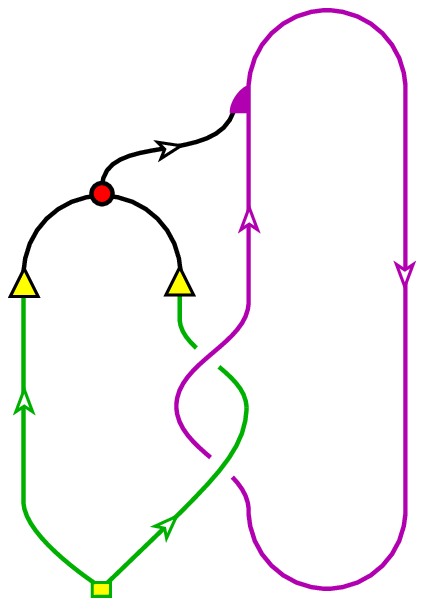}} }
    \put(0,0){
     \setlength{\unitlength}{.7pt}\put(-9,-322){
     \put( 47,456)   {\scriptsize$ A $}
     \put( 35,428)   {\scriptsize$ m $}
     \put( -2,414)   {\scriptsize$ e_W $}
     \put( 43,414)   {\scriptsize$ e_W $}
     \put( -2,353)   {\scriptsize$ U_W $}
     \put( 42,348)   {\scriptsize$ U_W $}
     \put( 87,417)   {\scriptsize$ M $}
     \put( 87,465)   {\scriptsize$ \rho_M $}
     }\setlength{\unitlength}{1pt}}
  \end{picture}} ~~.
\labl{eq:C-WM-graph}
By $\dim(\,\cdot\,)$ we denote the quantum dimension of an
object in $\Cc$.
The constants \erf{eq:C-WM-graph} have
the property $C(\one;A)=1$, as they should have
for \erf{eq:tft-bndstate} to hold (we take the Ishibashi
state $|\one\rrangle$ to be normalised as
$\langle 0|\one \rrangle=1$). Inserting \erf{eq:tft-bndstate}
into the definition \erf{eq:omegab-def} of $\X$ gives, for the
boundary condition $B_M$,
\be
  \X_M = \frac{2 \langle 0| W_2 \ol W{}_2 |U_W \rrangle}{c_1 c_2(c_1{+}c_2)}
  \cdot  
  \frac{\langle 0| A \rangle}{\langle 0| M \rangle }
  \cdot  
  C(U_W;M) ~~.
\labl{eq:X_M-aux1}
Since we know that $A$ corresponds to a factorising boundary
condition, we conclude
\be
  1 = \X_A = 
  \frac{2 \langle 0| W_2 \ol W{}_2 |U_W \rrangle}{c_1 c_2(c_1{+}c_2)}
  \cdot  
  C(U_W;A) ~~.
\ee
This, together with $\langle 0| A \rangle / \langle 0| M \rangle
= \dim(A)/\dim(M)$ (which follows as a special case
{}from \cite[sect.\,4.3]{Fuchs:2004xi}) allows one to simplify 
\erf{eq:X_M-aux1} to
\be
  \X_M = 
  \frac{\dim(A)}{\dim(M) }
  \cdot  
  \frac{C(U_W;M)}{C(U_W;A)} ~~.
\labl{eq:XM-via-C}
This expression is now defined entirely in terms of data
accessible on the level of the category $\Cc$ (as opposed
to \erf{eq:X_M-aux1} which e.g.\ makes explicit mention
of the central charges).
If furthermore $M$ is an induced module $\inda(V)$ for
some object $V$, then it is not too difficult to see 
that \erf{eq:XM-via-C} simplifies even further, namely to
\be
  \X_{\inda(V)} = \frac{s_{U_W,V}}{\dim(W)\dim(V)} ~~,
\labl{eq:X-IndA}
where $s_{U,V}$ denotes the invariant of the Hopf-link.
It is related to the modular $S$-matrix occurring in the
transformation of characters via $s_{U_i,U_j} = S_{ij}/S_{00}$
for two simple objects $U_i,U_j$. The quantum dimension
in turn can be expressed as $\dim(U_i) = S_{i0}/S_{00}$.

\subsubsection{Action of topological defects}\label{app:top-act}

The properties of topological defects, and how they can be
calculated in the TFT approach, has been discussed in detail
in \cite{Frohlich:2006ch}. Bimodules of $A$ describe topological
defects, and simple bimodules correspond to elementary topological defects.
As described in section \ref{sec:ref-trans}, topological defects 
can be fused. In terms of the $A$-bimodules, the corresponding
operation is the tensor product over $A$. That is, if $Y$, $Z$ are
$A$-bimodules, and $X_Y$, $X_Z$ are the corresponding defects,
then $X_Y \star X_Z = X_{Y \ota Z}$. The same holds for the
action on boundary conditions. Given an $A$-module $M$, one obtains
$X_Y \star B_M = B_{Y \ota M}$.

For models with charge conjugation modular invariant, like the
Lee-Yang, Ising, and $M_{2,7}$ minimal models treated
below, conformal boundary conditions and topological
defects are labelled by the irreducible representations
and their fusion is just given by the 
fusion product of the corresponding 
representations \cite{Petkova:2000ip,Petkova:2001ag}.
In the present context that follows since the relevant algebra
in this case is simply $A = \one$.

On the other hand, to understand the defect action for the 
$D_6$, $E_6$ and $E_8$-type modular invariants appearing below,
we need to evaluate the tensor products over the relevant
algebra $A$. To this end we need two additional concepts:
$\alpha$-induced bimodules and a procedure to turn a left module
into a bimodule. Given an object $U$ of $\Cc$, there are two natural
ways to define a bimodule structure on the object $A \otimes U$. Just
as for induced left modules, for the left action we simply use the
multiplication on $A$. For the right action, we need to take the object
$A$ past $U$ before we can use the multiplication. Since $\Cc$ is 
braided, there are two ways to do so (via over and under braiding), 
and we denote the resulting bimodules by $\alpha^+_A(U)$ and
$\alpha^-_A(U)$ 
(see e.g.\ \cite[def.\,2.21]{Frohlich:2003hm} for more details and 
references). If the two bimodules happen
to be isomorphic (as will be the case for the $\alpha$-induced bimodules
considered below), we just write $\alpha_A(U)$.
In the present application, the algebra $A$ is commutative. In this
case, every left module $M$ carries two natural bimodules structures.
The right action is defined by first using the over or under braiding
to take 
$A$ past $M$, and then applying the left representation morphism $\rho_M$
(in verifying that this is a right action one needs $A$ to be commutative).
If the module $M$ is local, it is easy to see that these two bimodule
structures coincide, and we will denote the resulting bimodule as
$M^\text{bi}$. One can check that, if $A$ is commutative and
the induced module $\inda(U)$ is local, then 
$\inda(U)^\text{bi} = \alpha_A(U)$. To compute the fusion of defects and
boundary conditions, the following rule is useful,
\be
  \alpha_A^\pm(U) \ota \inda(V) ~\cong~ \inda(U \oti V) ~~,
\ee
where $U$ and $V$ are objects of $\Cc$.
Writing out the definitions of $\alpha_A^\pm(U)$ and $\inda(V)$,
it is straightforward to give
an explicit isomorphism in terms
of the braiding of $\Cc$.

\subsection{Lee-Yang $\times$ Lee-Yang}\label{app:LYxLY}

The product of two $M_{2,5}$ minimal models can be described
as the $D_6$ invariant of $M_{3,10}$. The Kac tables of these
minimal models are:

\vspace{-.3em}
\begin{center}
\begin{tabular}{c||c|c|c|c|}
\multicolumn{5}{c}{$M_{2,5}$} \\[.2em]
\hline 
& \hspace*{1.5em} & \hspace*{1.2em} & \hspace*{1.2em} &
\hspace*{1.5em}\\[-.6em]
1 & 0 & $\!{-}\tfrac15\!$ & $\!{-}\tfrac15\!$ & 0 \\[0.3em]
\hline\hline 
 &   &   &   &   \\[-.8em]
 & 1 & 2 & 3 & 4
\end{tabular}
\hspace*{3em}
\begin{tabular}{c||c|c|c|c|c|c|c|c|c|}
\multicolumn{10}{c}{$M_{3,10}$} \\[.2em]
\hline 
& \hspace*{1.2em} & \hspace*{1.2em} & \hspace*{1.2em} &
\hspace*{1.2em} & \hspace*{1.2em} & \hspace*{1.2em} & \hspace*{1.2em} 
& \hspace*{1.2em} & \hspace*{1.2em}  \\[-.6em]
2 & $2$ & $\tfrac{49}{40}$ & $\tfrac{3}{5}$ & $\tfrac{1}{8}$ & 
    $\!{-}\tfrac{1}{5}\!$ & $\!{-}\tfrac{3}{8}\!$ &
    $\!{-}\tfrac{2}{5}\!$ 
&
    $\!\!{-}\tfrac{11}{40}\!\!$ & $0$ \\[.3em]
\hline
 &   &   &   &   &   &   &   &   &   \\[-.6em]
1 & $0$ & $\!\!{-}\tfrac{11}{40}\!\!$ & $\!{-}\tfrac{2}{5}\!$ 
  & $\!{-}\tfrac{3}{8}\!$ & 
    $\!{-}\tfrac{1}{5}\!$ & $\tfrac{1}{8}$ & $\tfrac{3}{5}$ &
    $\tfrac{49}{40}$ & $2$ \\[.3em]
\hline\hline 
 &   &   &   &   &   &   &   &   &   \\[-.8em]
 & 1 & 2 & 3 & 4 & 5 & 6 & 7 & 8 & 9  
\end{tabular}
\end{center}

Let us denote the irreducible representation of $M_{3,10}$ with 
Kac label $(r,s)$ by $U_{r,s}$. Then the algebra $A$ is given
by $A = U_{1,1} \oplus U_{1,9}$ (as an object, we do not need
here the explicit form of the multiplication), and $W$ is the primary
state in the representation $U_W = U_{1,9}$.

By the method of induced modules and \erf{eq:induced-homs} one finds
that 
$\inda(U_{1,1})$, $\inda(U_{1,2})$, $\inda(U_{1,3})$, $\inda(U_{1,4})$
are simple and that $\inda(U_{1,5}) = M_+ \oplus M_-$ for two simple
$A$-modules $M_{\pm}$. These are all simple modules of $A$. 
As objects they decompose as
\bea
  \inda(U_{1,1}) = U_{1,1} \oplus U_{1,9} ~~,\quad
  \inda(U_{1,2}) = U_{1,2} \oplus U_{1,8} ~~,\quad
  \\[.5em]
  \inda(U_{1,3}) = U_{1,3} \oplus U_{1,7} ~~,\quad
  \inda(U_{1,4}) = U_{1,4} \oplus U_{1,6} ~~,\quad
  M_{\pm} = U_{1,5} ~~.
\eear\ee
A glance at the conformal weights mod $\Zb$ reveals that
$\inda(U_{1,1})$, $\inda(U_{1,3})$ and $M_\pm$ are local,
and, denoting the representations of $M_{2,5}$ by 
$\one$ (for weight 0) and $\phi$ (for weight ${-}\tfrac15$), that
in terms of the product model
\be
  \inda(U_{1,1}) = \one \times \one ~~,\quad 
  \inda(U_{1,3}) = \phi \times \phi ~~,\quad
  M_+ = \phi \times \one ~~,\quad
  M_- = \one \times \phi ~~,
\labl{eq:M25-fields}
where the last two identifications are a choice of convention.

To see the $D_6$ diagram appear we have to draw a dot for each
boundary condition and draw a line between two dots whenever
the representation $U_{1,2}$ appears in the space of states
for the strip with these two boundary conditions on either side
\cite{Behrend:1998fd,Behrend:1999bn}. In the present language
this means we draw a dot for each $A$-module, and given
two $A$-modules $M, N$ we draw 
$\dim\Hom_A(M \oti U_{1,2},N)$ lines between them.
Using that for induced modules $\inda(U) \oti V = \inda(U \oti V)$
one finds
\be
  \raisebox{-30pt}{
  \begin{picture}(200,60)
    \put(0,5){ \scalebox{.8}{\includegraphics{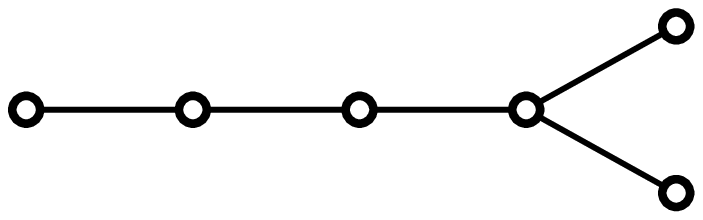}} }
    \put(0,5){
     \setlength{\unitlength}{.8pt}\put(-59,-46){
     \put( 60,60)   {\scriptsize$ A $}
     \put( 85,60)   {\scriptsize$ \inda(U_{1,2}) $}
     \put(145,60)   {\scriptsize$ \inda(U_{1,3}) $}
     \put(180,90)   {\scriptsize$ \inda(U_{1,4}) $}
     \put(266,97)   {\scriptsize$ M_+ $}
     \put(266,49)   {\scriptsize$ M_- $}
     }\setlength{\unitlength}{1pt}}
  \end{picture}}
\labl{eq:D6-app}
For example, $\dim\Hom_A(\inda(U_{1,4}) \oti U_{1,2} , M_+)
= \dim\Hom(U_{1,4} \oti U_{1,2} , U_{1,5}) = 1$.

We know that the product model has four factorising boundary
conditions 
and two permutation boundary conditions. 
This already accounts for all six conformal
boundary conditions of the $D$-invariant of $M_{3,10}$.
The local modules give factorising boundary conditions, so that
the two permutation ones correspond to
the boundary conditions $B_{\inda(U_{1,2})}$ and 
$B_{\inda(U_{1,4})}$.
To identify which is which, we compute
the action of the topological defect $X_\phi \times X_\phi$ on 
the boundary condition $B_{\inda(U_{1,2})}$. Expressing this in terms of
bimodules, one finds
$\alpha_A(U_{1,3}) \ota \inda(U_{1,2}) \cong 
\inda(U_{1,2}) \oplus \inda(U_{1,4})$. The fact that this is a direct
sum of two simple modules, rather than four, is only consistent
with 
\be
  B_{\inda(U_{1,2})} = \text{folded~defect}~X_\one \quad , \quad  
  B_{\inda(U_{1,4})} = \text{folded~defect}~X_\phi ~~.
\labl{eq:M25-defect}
The action of the topological defects of $M_{2,5} \times M_{2,5}$
on all boundary conditions listed in \erf{eq:D6-app} is now simply
computed by noting that for $A$-series models the fusion of
a defect with another defect or a boundary condition agrees with
the fusion of the chiral representations labelling them.
{}From the identifications \erf{eq:M25-fields} and \erf{eq:M25-defect}
we then conclude, for example,
$(X_\phi \times X_\one) \star B_{M_-} =  B_{\inda(U_{1,3})}$
or 
$(X_\one \times X_\phi) \star B_{\inda(U_{1,4})} =  B_{\inda(U_{1,2})}
+ B_{\inda(U_{1,4})}$.
In this way one can obtain all the data presented in \erf{eq:M25xM25-D6}.

\subsection{Lee-Yang $\times$ Ising}\label{app:LYxIs}

The product $M_{2,5} \times M_{3,4}$ is equivalent to the $E_6$-invariant
of $M_{5,12}$. The Kac table of $M_{2,5}$ was given in the previous
section, 
and those of $M_{3,4}$ and $M_{5,12}$ are:

\begin{center}
\begin{tabular}{c||c|c|c|}
\multicolumn{4}{c}{$M_{3,4}$} \\[.2em]
\hline 
& \hspace*{1.5em} & \hspace*{1.2em} & \hspace*{1.2em} \\[-.6em]
2 & $\tfrac12$ & $\!\tfrac{1}{16}\!$ & $0$  \\[0.3em]
\hline
 &   &   &    \\[-.8em]
1 & $0$ & $\!\tfrac{1}{16}\!$ & $\tfrac12$  \\[0.3em]
\hline\hline 
 &   &   &    \\[-.8em]
 & 1 & 2 & 3 
\end{tabular}
\hspace*{2em}
\begin{tabular}{c||c|c|c|c|c|c|c|c|c|c|c|}
\multicolumn{12}{c}{$M_{5,12}$} \\[.2em]
\hline 
& \hspace*{1.2em} & \hspace*{1.2em} & \hspace*{1.2em} &
\hspace*{1.2em} & \hspace*{1.2em} & \hspace*{1.2em} & \hspace*{1.2em} 
& \hspace*{1.2em} & \hspace*{1.2em} & \hspace*{1.2em} &
\hspace*{1.2em}  
\\[-.6em]
4 & $\tfrac{15}{2}$ & $\tfrac{93}{16}$ & $\tfrac{13}{3}$ 
  & $\tfrac{49}{16}$ & 
    $2$ & $\tfrac{55}{48}$ & $\tfrac{1}{2}$ &
    $\tfrac{1}{16}$ & $\!{-}\tfrac{1}{6}\!$ &
    $\!\!{-}\tfrac{3}{16}\!\!$ 
& $0$ \\[.3em]
\hline
 &   &   &   &   &   &   &   &   &   &    & \\[-.8em]
3 & $\tfrac{19}{5}$ & $\!\tfrac{209}{80}\!\!$ & $\tfrac{49}{30}$ 
  & $\tfrac{69}{80}$ & 
    $\tfrac{3}{10}$ & $\!\!\tfrac{{-}13}{240}\!\!$ &
    $\!\!{-}\tfrac{1}{5}$ 
&
    $\!\!{-}\tfrac{11}{80}\!\!$ & $\tfrac{2}{15}\!$ & 
    $\tfrac{49}{80}$ & $\tfrac{13}{10}$ \\[.3em]
\hline
 &   &   &   &   &   &   &   &   &   &    & \\[-.8em]
2 & $\tfrac{13}{10}$ & $\tfrac{49}{80}$ & $\tfrac{2}{15}\!$ 
  & $\!\!{-}\tfrac{11}{80}\!\!$ & 
    $\!\!{-}\tfrac{1}{5}$ & $\!\!\tfrac{{-}13}{240}\!\!$ 
& $\tfrac{3}{10}$ &
    $\tfrac{69}{80}$ & $\tfrac{49}{30}$ & 
    $\!\tfrac{209}{80}\!\!$ & $\tfrac{19}{5}$ \\[.3em]
\hline
 &   &   &   &   &   &   &   &   &   &    & \\[-.8em]
1 & $0$ & $\!\!{-}\tfrac{3}{16}\!\!$ & $\!{-}\tfrac{1}{6}\!$ 
  & $\tfrac{1}{16}$ & 
    $\tfrac{1}{2}$ & $\tfrac{55}{48}$ & $2$ &
    $\tfrac{49}{16}$ & $\tfrac{13}{3}$ & $\tfrac{93}{16}$ 
& $\tfrac{15}{2}$ \\[.3em]
\hline\hline 
 &   &   &   &   &   &   &   &   &   &    & \\[-.8em]
 & 1 & 2 & 3 & 4 & 5 & 6 & 7 & 8 & 9 & 10 & 11
\end{tabular}
\end{center}

\noindent
To have a unique labelling of the irreducible representations of
$M_{5,12}$ 
we restrict to the range
of Kac labels $(r,s)$ for which $r \in \{1,3\}$. Also, as before, by
$\one$ and 
$\phi$ we will denote the two irreducible representations for $M_{2,5}$.
For the Ising model $M_{3,4}$ we choose the labels $\one$, $\sigma$ and
$\eps$ for the representations of highest weight $0$, $\tfrac1{16}$ and
$\tfrac12$, respectively.

The primary field $W$ is the highest weight state of the representation
$U_W = U_{1,7}$, and the relevant commutative Frobenius
algebra is built on the object $A = U_{1,1} \oplus U_{1,7}$. Next we
work out how the induced $A$-modules decompose into simple modules. For
$r \in \{1,3\}$, the result
is
\be\begin{array}{ll}\displaystyle
\inda(U_{r,1}) ~~\text{(simple)}
\etb
\inda(U_{r,7}) \cong \inda(U_{r,1}) \oplus \inda(U_{r,3})
\enl
\inda(U_{r,2}) ~~\text{(simple)}
\etb
\inda(U_{r,8}) \cong \inda(U_{r,2}) \oplus M_r
\enl
\inda(U_{r,3}) ~~\text{(simple)}
\etb
\inda(U_{r,9}) \cong \inda(U_{r,3})
\enl
\inda(U_{r,4}) \cong \inda(U_{r,10}) \oplus M_r
\etb
\inda(U_{r,10}) ~~\text{(simple)}
\enl
\inda(U_{r,5}) \cong \inda(U_{r,3}) \oplus \inda(U_{r,11}) \quad
\etb
\inda(U_{r,11}) ~~\text{(simple)}
\enl
\inda(U_{r,6}) \cong \inda(U_{r,2}) \oplus \inda(U_{r,10})
\end{array}\ee
For each value of $r \in \{1,3\}$ there are six simple modules: the
induced 
modules for the objects $U_{r,1}$, $U_{r,2}$, $U_{r,3}$, $U_{r,10}$,
$U_{r,11}$, 
and the module $M_r$. The Dynkin diagram is again found by computing
$\dim\Hom_A( M \oti U_{1,2} , N)$ for each pair of simple modules $M$,
$N$. 
This gives the $E_6$ diagram, as expected,
\be
  \raisebox{-35pt}{
  \begin{picture}(200,70)
    \put(0,15){ \scalebox{.8}{\includegraphics{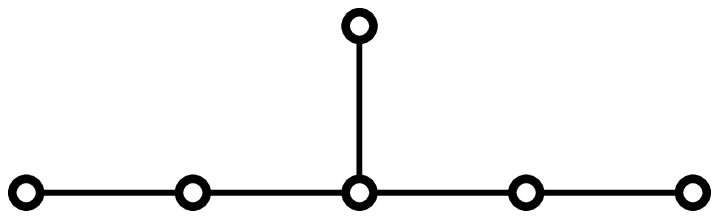}} }
    \put(0,15){
     \setlength{\unitlength}{.8pt}\put(-59,-70){
     \put( 45,60)   {\scriptsize$ \inda(U_{r,1}) $}
     \put( 95,88)   {\scriptsize$ \inda(U_{r,2}) $}
     \put(145,60)   {\scriptsize$ \inda(U_{r,3}) $}
     \put(195,88)   {\scriptsize$ \inda(U_{r,10}) $}
     \put(245,60)   {\scriptsize$ \inda(U_{r,11}) $}
     \put(173,123)   {\scriptsize$ M_r $}
     }\setlength{\unitlength}{1pt}}
  \end{picture}}
\labl{eq:E6-app}
It is also easy to see that the
simple $A$-modules consist of the following irreducible $M_{5,12}$
representations,
\be\begin{array}{ll}\displaystyle
\inda(U_{r,1}) = U_{r,1} \oplus U_{r,7}
\etb
\inda(U_{r,10}) = U_{r,4} \oplus U_{r,6} \oplus U_{r,10}
\enl
\inda(U_{r,2}) = U_{r,2} \oplus U_{r,6} \oplus U_{r,8}
\etb
\inda(U_{r,11}) = U_{r,5} \oplus U_{r,11}
\enl
\inda(U_{r,3}) = U_{r,3} \oplus U_{r,5} \oplus U_{r,7} \oplus U_{r,9}
\qquad\etb
M_r = U_{r,4} \oplus U_{r,8}
\end{array}\ee
The weights mod $\Zb$ of these irreducible representations as given
in the Kac table above then tell us which of these simple modules
are local, and how they are identified with the irreducible
representations 
of $M_{2,5} \times M_{3,4}$. One finds the following local, simple
$A$-modules
\be\begin{array}{lll}\displaystyle
  \inda(U_{1,1}) = \one \times \one \quad\etb
  M_1 = \one \times \sigma  \quad\etb
  \inda(U_{1,11}) = \one \times \eps  
  \enl
  \inda(U_{3,1}) = \phi \times \one  \quad\etb
  M_3 = \phi \times \sigma  \quad\etb
  \inda(U_{3,11}) = \phi \times \eps\;\;.
\eear\labl{eq:M512-localmod}
Next we turn to the action of topological defects 
in each of the two
factors on the boundary conditions \erf{eq:E6-app}, i.e.\ the action
of $X_\phi \times X_\one$, $X_\one \times X_\sigma$ and $X_\one \times
X_\eps$. 
For $X_\phi \times X_\one$, this action amounts to tensoring with the
bimodule $\alpha_A(U_{3,1})$, and one simply gets, for $N_r = 
\inda(U_{r,s})$ or $N_r = M_r$,
\be
  \alpha_A(U_{3,1}) \ota N_1 = N_3 \quad , \qquad
  \alpha_A(U_{3,1}) \ota N_3 = N_1 \oplus N_3  ~~.
\ee
This explains the effect on the odd nodes of the $A_4$ diagram in
\erf{eq:Ly-Is-nodes}. The action of $X_\one \times X_\eps$ is equally
easy to obtain. First of all,
$\alpha_A(U_{1,11}) \ota \inda(U_{r,s}) = \inda(U_{r,12-s})$.
The action on $M_r$ can then be found by applying
$\alpha_A(U_{1,11}) \ota(\, - \,)$ to both sides
of $\inda(U_{r,4}) \cong \inda(U_{r,10}) \oplus M_r$, which
then gives $\alpha_A(U_{1,11}) \ota M_r = M_r$.

For the action of $M_1^\text{bi}$ (the bimodule obtained from the
local module $M_1$ as in section \ref{app:top-act})
one can compare the action of
$\inda(U_{1,4})$ and $\inda(U_{1,10})$. For example,
\be\begin{array}{ll}
  \alpha_A(U_{1,4}) \,\,\ota \inda(U_{r,2})
  \etb \cong ~\inda(U_{r,11}) \oplus \inda(U_{r,3}) \oplus
  \inda(U_{r,3}) 
  \enl
  \alpha_A(U_{1,10}) \ota \inda(U_{r,2})
  \etb \cong ~\inda(U_{r,11}) \oplus \inda(U_{r,3})
\eear\ee
{}from which we conclude that $M_1^\text{bi} \ota \inda(U_{1,2}) \cong 
\inda(U_{1,3})$. 

So far we have outlined how to obtain the data presented in
figure \erf{eq:Ly-Is-nodes}. Let us now turn to the computation of
the coefficients $\X$. For the factorising boundary conditions,
i.e.\ those corresponding to the modules \erf{eq:M512-localmod},
one has $\X=1$. The remaining boundary conditions are related by
the action of the topological defects, and hence it is enough
to compute $\X$ for one representative. We pick the
module $\inda(U_{1,2})$, because in this case the simplified
formula \erf{eq:X-IndA} is applicable. To evaluate it, we need
\be
  \dim U_{1,2} = - \frac{\sqrt{2}}{1+\sqrt{3}} ~~,\quad
  \dim U_{1,7} = 2 - \sqrt{3}  ~~,\quad
  s_{U_{1,2},U_{1,7}} = \frac{\sqrt{2}}{1+\sqrt{3}}  ~~.
\ee
The result is $\X_{\inda(U_{1,2})} = {-}2-\sqrt{3}$. Inserting this
value into \erf{eq:RT-via-omega} gives the reflection and transmission  
coefficients stated in \erf{eq:M25-M34-refl-trans}.

\subsection{Lee-Yang $\times \,{\bf M_{2,7}}$}\label{app:LYxM27}

The final example of a product of minimal models that can be
described as an extension of another minimal model is
$M_{2,5} \times M_{2,7}$, which is equivalent to the 
$E_8$-invariant of $M_{7,30}$. The minimal model $M_{2,7}$ has
central charge $c={-}\tfrac{68}7$ and Kac table

\begin{center}
\begin{tabular}{c||c|c|c|c|c|c|}
\multicolumn{7}{c}{$M_{2,7}$} \\[.2em]
\hline 
& \hspace*{1.5em} & \hspace*{1.2em} & \hspace*{1.2em} 
& \hspace*{1.5em}& \hspace*{1.5em}& \hspace*{1.5em}\\[-.6em]
1 & 0 & $\!{-}\tfrac27\!$ & $\!{-}\tfrac37\!$ 
& $\!{-}\tfrac37\!$ & $\!{-}\tfrac27\!$ & 0 \\[0.3em]
\hline\hline 
 &   &   &   &   &   &   \\[-.8em]
 & 1 & 2 & 3 & 4 & 5 & 6
\end{tabular}
\end{center}

\noindent
Let us denote the irreducible representation of weight ${-}\tfrac27$
by $\alpha$ and that of weight ${-}\tfrac37$ by $\beta$. The fusion
rules are then $\alpha \oti \alpha = \one \oplus \beta$,
$\alpha \oti \beta = \alpha \oplus \beta$ and $\beta \oti \beta = 
\one \oplus \alpha \oplus \beta$.

We will not state the Kac table of $M_{7,30}$ explicitly. The
relevant algebra is built on the representation
$A = U_{1,1} \oplus U_{1,11} \oplus U_{1,19} \oplus U_{1,29}$
whose irreducible summands have highest weights 
$0$, $2$, $12$ and $35$, respectively. In particular, the
chiral field $W$ is a highest weight state in 
$U_W = U_{1,11}$. Note that $J \equiv U_{1,29}$ is a simple
current which acts under fusion as $J \oti U_{r,s} \cong U_{r,30-s}$.
Since $A \oti J \cong A$ it is not hard to convince oneself that
for induced modules,
\be
  \inda(U) \cong \inda(J \oti U) ~~.
\ee
It is thus enough to consider the induced modules for
$U_{r,s}$ with $s \in \{1,2,\dots,15\}$. These induced modules decompose
as follows into simple modules
\be\begin{array}{ll}\displaystyle
\inda(U_{r,1}) ~~\text{(simple)}
  \etb
\inda(U_{r,9})\,\, \cong \inda(U_{r,3}) \oplus \inda(U_{r,5})
\enl
\inda(U_{r,2}) ~~\text{(simple)}
  \etb
\inda(U_{r,10}) \cong \inda(U_{r,2}) \oplus \inda(U_{r,4}) 
\oplus M^{(2)}_r
\enl
\inda(U_{r,3}) ~~\text{(simple)}
  \etb
\inda(U_{r,11}) \cong \inda(U_{r,1}) \oplus \inda(U_{r,3}) 
\oplus \inda(U_{r,5})
\enl
\inda(U_{r,4}) ~~\text{(simple)}
  \etb
\inda(U_{r,12}) \cong  \inda(U_{r,2}) \oplus \inda(U_{r,4}) 
\oplus M^{(1)}_r
\enl
\inda(U_{r,5}) ~~\text{(simple)}
  \etb
\inda(U_{r,13}) \cong \inda(U_{r,3}) \oplus \inda(U_{r,5}) 
\oplus M^{\phi}_r
\enl
\inda(U_{r,6}) \cong M_r^{(1)} \oplus M_r^{(2)}
  \etb
\inda(U_{r,14}) \cong \inda(U_{r,4}) \oplus M_r^{(1)} \oplus M_r^{(2)}
\enl
\inda(U_{r,7}) \cong \inda(U_{r,5}) \oplus M_r^{\phi} \quad
  \etb
\inda(U_{r,15}) \cong 2\,\inda(U_{r,5})
\enl
\inda(U_{r,8}) \cong \inda(U_{r,4}) \oplus M_r^{(1)}
\end{array}\labl{eq:M730-decomp}
Here $r$ takes values in $\{1,3,5\}$. There are thus 24 simple
$A$-modules, 
namely $\inda(U_{r,s})$ with $s \in \{1,2,3,4,5\}$, as well as
$M^{(1)}_r$, $M^{(2)}_r$ and $M^{\phi}_r$.
As before one can find the decomposition of these simple modules
into representations of $M_{7,30}$, and by evaluating the conformal
weight modulo $\Zb$ one finds that there are six local modules.
Their identification with representations of the product model is
as follows
\be\begin{array}{lll}\displaystyle
  \inda(U_{1,1}) = \one   \times \one \quad\etb
  \inda(U_{3,1}) = \beta  \times \one  \quad\etb
  \inda(U_{5,1}) = \alpha \times \one  
  \enl
  M_1^{\phi} = \one   \times \phi \quad\etb
  M_3^{\phi} = \beta  \times \phi \quad\etb
  M_5^{\phi} = \alpha \times \phi
\eear\labl{eq:M730-localmod}
To obtain the $E_8$ diagram we need to compute the effect of tensoring
the simple modules with $U_{1,2}$ from the right. 
For induced modules this is again an easy exercise. For the
simple modules not isomorphic to induced modules one can, for example,
tensor both sides of $\inda(U_{r,7}) \cong \inda(U_{r,5}) \oplus
M_r^{\phi}$ from the right with $U_{1,2}$ which gives
\be
  \inda(U_{r,6}) \oplus   \inda(U_{r,8}) ~\cong~
  \inda(U_{r,4}) \oplus   \inda(U_{r,6}) \oplus (M_r^{\phi} \oti U_{1,2}) ~~.
\ee
Substituting the decompositions in \erf{eq:M730-decomp} it follows
that $M_r^{\phi} \oti U_{1,2} \cong M_r^{(1)}$. For $M_r^{(1)}$
and $M_r^{(2)}$ one can proceed similarly. Altogether one finds,
for $r \in \{1,3,5\}$,
\be
  \raisebox{-35pt}{
  \begin{picture}(300,70)
    \put(0,15){ \scalebox{.8}{\includegraphics{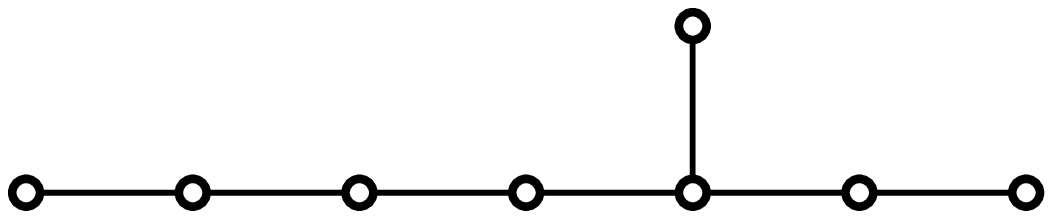}} }
    \put(0,15){
     \setlength{\unitlength}{.8pt}\put(-59,-70){
     \put( 45, 60)   {\scriptsize$ \inda(U_{r,1}) $}
     \put( 95, 88)   {\scriptsize$ \inda(U_{r,2}) $}
     \put(145, 60)   {\scriptsize$ \inda(U_{r,3}) $}
     \put(195, 88)   {\scriptsize$ \inda(U_{r,4}) $}
     \put(245, 60)   {\scriptsize$ \inda(U_{r,5}) $}
     \put(305, 88)   {\scriptsize$ M_r^{(1)} $}
     \put(355, 60)   {\scriptsize$ M_r^{\phi} $}
     \put(270,123)   {\scriptsize$ M_r^{(2)} $}
     }\setlength{\unitlength}{1pt}}
  \end{picture}}
\labl{eq:E8-app}
The action of the topological defects $X_\one \times X_\alpha$, 
$X_\one \times X_\beta$ and $X_\phi \times \one$ on these conformal
boundary conditions corresponds to tensoring the corresponding
$A$-modules over $A$ from the left with
$\alpha_A(U_{5,1})$,
$\alpha_A(U_{3,1})$, and
$M_1^{\phi,\text{bi}}$, respectively. To compute these tensor products,
it is easiest to
work with isomorphism classes rather than directly with modules and
bimodules. In this way we obtain the fusion ring of bimodules, and a
representation 
of that ring on the $\Zb$-module generated by the isomorphism classes
of simple modules. 
The point is that in this setting it makes sense to consider differences.
For a bimodule $B$ or a module $M$, denote by 
$[B]$ and $[M]$ the corresponding isomorphism classes. Then
\bea
  [M_1^{\phi,\text{bi}}] = [\alpha_A(U_{1,7})] - [\alpha_A(U_{1,5})] ~~,
\enl
  [M_r^{(1)}] = [\inda(U_{1,8})] - [\inda(U_{1,4})]
  ~~,\quad
  [M_r^{\phi}] = [\inda(U_{1,7})] - [\inda(U_{1,5})]~,
\enl
  [M_r^{(2)}] = [\inda(U_{1,6})] - [\inda(U_{1,8})] + [\inda(U_{1,4})]~.
\eear\ee
One can now compute, for example,
\bea
  [ M_1^{\phi,\text{bi}} \ota \inda(U_{r,2}) ]
  = [ M_1^{\phi,\text{bi}} ] . [ \inda(U_{r,2}) ]
\enl
  = [\alpha_A(U_{1,7})].[\inda(U_{r,2})] 
    - [\alpha_A(U_{1,5})].[\inda(U_{r,2})]
\enl
  = [\alpha_A(U_{1,7}) \ota \inda(U_{r,2})] - 
    [\alpha_A(U_{1,5}) \ota \inda(U_{r,2})]
\enl
  = [\inda(U_{r,6}) \oplus \inda(U_{r,8})] - 
    [\inda(U_{r,4}) \oplus \inda(U_{r,6})]
\enl
  = [\inda(U_{r,6})] + [\inda(U_{r,8})] - 
    [\inda(U_{r,4})] - [\inda(U_{r,6})]
  = [M_r^{(1)}]
\eear\ee
i.e.\ $M_1^{\phi,\text{bi}} \ota \inda(U_{r,2})  \cong M_r^{(1)}$.
Proceeding along these lines, one finds the action of the topological
defects as given in \erf{eq:M25xM27-E8}.

Finally, we need to calculate the coefficients $\X$ for the various
conformal defects corresponding to the folded boundary conditions of
the folded model. Due to the action of topological defects as
given in \erf{eq:M25xM27-E8}, by comparing to \erf{eq:E8-app} we
see that it is enough to compute $\X$ for the representatives
$\inda(U_{1,s})$ with $s \in \{1,2,3,4\}$.
Since all of these are induced modules, we can again apply
\erf{eq:X-IndA}. The following constants are needed,
\be\bearll
  \dim(U_{1,s}) \etb\!\!=\, (-1)^{s-1} 
  \sin\!\big(\pi \tfrac{7s}{30} \big) 
/ \sin\big(\pi \tfrac{7}{30} \big) \;,
  \enl
  s_{U_W,U_{1,s}} \etb\!\!=\, (-1)^{s-1} 
  \sin\!\big(\pi\tfrac{77s}{30} \big) 
/ \sin\big(\pi \tfrac{7}{30} \big) \,.
\eear\ee
Substituting these into \erf{eq:X-IndA} one recovers 
\erf{eq:M27-Xb-values}.

\raggedright
\end{document}